\documentclass[fleqn,usenatbib]{mnras}

\usepackage{newtxtext,newtxmath}
\usepackage[T1]{fontenc}

\DeclareRobustCommand{\VAN}[3]{#2}
\let\VANthebibliography\thebibliography
\def\thebibliography{\DeclareRobustCommand{\VAN}[3]{##3}\VANthebibliography}

\usepackage{graphicx}	% Including figure files
\usepackage{amsmath}	% Advanced maths commands

\usepackage{float}

\title[Machine vision and galaxy structure]{Deciphering galaxy images using machine vision -- Combining variational autoencoder and principal component analysis for feature extraction}
\author[S. Howie et al.]{Samuel Howie,$^{1}$
Ting-Yun Cheng,$^{2,3}$%\thanks{E-mail:xxx}
Carlton M. Baugh,$^{4}$
\\ \\
$^{1}$ Department of Physics, Durham University, South Road, Durham, DH1 3LE, UK.\\
$^{2}$ Centre for Extragalactic Astronomy, Department of Physics, Durham University, South Road, Durham, DH1 3LE, UK\\
$^{3}$ Kapteyn Astronomical Institute, University of Groningen, Landleven 12 (Kapteynborg, 5419), 9747 AD Groningen, The Netherlands\\
$^{4}$ Institute for Computational Cosmology, Departmnent of Physics, Durham University, South Road, Durham, DH1 3LE, UK.}

\date{Accepted XXX. Received YYY; in original form ZZZ}

\pubyear{\the\year{}}

\begin{document}
\label{firstpage}
\pagerange{\pageref{firstpage}--\pageref{lastpage}}
\maketitle

% Abstract of the paper
\begin{abstract}
Understanding the visual features contained in galaxy images is of utmost importance for interpreting their underlying structures. Here, we present a machine vision approach, combining a variational autoencoder (VAE) framework with principal component analysis (PCA), to decipher galaxy images. Using mock $gri$-band  images from the EAGLE simulation, the VAE finds that around 35 features are needed to describe the images. Adding the PCA, we identify an optimal range of 10-12 features needed to capture 99.9 per cent of the variance in galaxy images. The exact optimal number varies with galaxy complexity: disk-dominated galaxies require 12 features, bulge-dominated galaxies need 9, and intermediate systems require 10-11 features.
Correlations between extracted PCA features and structural measurements reveal that the VAE prioritizes galaxy size during reconstruction, with half-light radius strongly correlating with the highest-ranked principal components. Subsequent features capture morphology-dependent characteristics: disk-dominated galaxies emphasize size, asymmetry, and position angle; bulge-dominated systems focus on size, concentration, and axis ratio; while intermediate galaxies show enhanced attention to S\'ersic index, indicating greater emphasis on accurately reproducing both disk and bulge structures.
The PCA process significantly reduces the entanglement of the features compared to the raw VAE latent features, decreasing the correlations with the half-light radius and the S\'ersic index from $14.5 \pm 1.0$ and $6.0 \pm 1.5$ features, respectively, to only $2.0 \pm 1.0$ components after PCA. Using Uniform Manifold Approximation and Projection (UMAP), we construct two-dimensional visualizations that preserve neighborhood relationships from the high-dimensional feature space. This demonstrates that machine vision can effectively distinguish galaxy populations across different morphological types, including systems with atypical structures that may be overlooked by traditional classification methods, providing a data-driven complement to conventional structural measurements.
\end{abstract}

% Select between one and six entries from the list of approved keywords.
% Don't make up new ones.

%%%%%%%%%%%%%%%%%%%%%%%%%%%%%%%%%%%%%%%%%%%%%%%%%%

%%%%%%%%%%%%%%%%% BODY OF PAPER %%%%%%%%%%%%%%%%%%

\section{Introduction}
\label{sec:intro}

    There are billions of galaxies in the Universe, with wide ranging morphological characteristics. The morphology of a galaxy is the most direct outcome of its evolutionary history, influenced by multiple effects and events, such as interactions with the environment and other galaxies, as well as transformations driven by internal physical processes (e.g., \citealt{Dressler1980,Whitmore1993,Goto2003,Kauffmann2003,Kormendy2004,Boselli2006,Park2007,VandeWel2008,Gadotti2011,Wilman2013,Bait2017,Rodriguez-Gomez2017,Thob2019,Bittner2020,Irodotou2022,Fragkoudi2024,Aritra2024,Ansar2025,Kawinwanichakij2025}). The structure of a galaxy is therefore closely linked to its physical properties, such as stellar population (in particular the age of the population and its metallicity), stellar mass, and star-formation activity. Thus, understanding the categorisation of galaxy morphologies and their origins is essential for unravelling the physical processes that drive galaxy formation and evolution. 
    
    Pioneering work, such as Hubble's proposal of a morphological sequence \citep{Hubble1926}, utilised visual inspection of galaxy images to categorise the diverse morphologies of galaxies into two main classes: elliptical and spiral galaxies. 
    %is the oldest and most widely accepted galaxy classification scheme grouping galaxies, based on appearance into elliptical and spiral galaxies. 
    These visual classes are backed up by other physical differences between these galaxies, such as star formation rate and stellar population age. 
    The Hubble classification scheme has been refined to include additional features, such as lenticular galaxies, barred spirals, ring structures, etc. \citep[e.g.][]{Hubble1936,Sandage1961,deVaucouleurs1959,Elmegreen1982,Elmegreen1987}, providing a more comprehensive framework for describing the diversity of galaxy morphologies. However, visual inspection is not only labourious, making it challenging to keep up with the rapidly growing volume of astronomical data from modern surveys, but also inherently subjective and prone to human bias \citep{Abraham2001}. %It is strongly influenced by the experience of the classifier and image quality . 
    Moreover, with larger wide-field surveys, such as the Sloan Digital Sky Survey \citep[SDSS;][]{York2000}, deviations from the commonly known relationships between visual galaxy morphology, colour, and star formation activities have been identified. For example, the discovery of red spiral galaxies challenges the traditional picture that spirals are blue and actively star-forming, while red galaxies are typically passive ellipticals \citep[][]{Masters2010}. 
    
    As a result, many studies have explored numerical and quantitative approaches, employing both parametric (e.g. Sersic profile) and nonparametric methods (e.g. concentration-asymmetry-smoothness parameters, Gini coefficient, M20, etc.) to analyse galaxy structures \citep[e.g.][]{Sersic1963,Sersic1968,Conselice2000,Bershady2000,Abraham2003,Conselice2003,Lotz2004,Graham2005,Law2007,Andrae2011}. These methods not only provide a more objective framework for inspecting the structure of galaxies, potentially mitigating subjective judgment and dependence on image quality, but also offer a scalable solution for handling the vast amount of survey data. These methods analyse the light distribution in an image and employ different mathematical formulations to emphasize specific features that might be associated with the structure of a galaxy. Although the combination of these measurements has led to many successful galaxy classifications, particularly at higher redshifts where image quality is lower \citep[e.g.][see also the review by \citealt{Conselice2014}]{Conselice2003,Lotz2004,Conselice2008,Bluck2012,Ferreira2022}, they still require prior human knowledge. Some of the methods struggle to achieve a clear decision boundary in classifying galaxies. Moreover, \citet{Andrae2011} demonstrated that the commonly used morphological parameters are intertwined and should be estimated independently. They also found that combining these parameters does not resolve this interdependence, highlighting the need for alternative approaches or new parameters to better capture the complexity of galaxy morphology \citep[e.g.][]{Ferrari2015,Rosa2018}. 

    This work has inspired the development of various approaches that apply machine learning (ML) techniques, particularly unsupervised ML (UML) methods, which do not require prior knowledge of the datasets. Early UML applications to galaxy morphology can be traced to \citet{Schutter2015}, who applied a set of image descriptors, such as texture analysis, polynomial decompositions, pixel statistics, Fourier and wavelet transformations, etc, to analyse galaxy images. More recent work in this area includes \citet{Hocking2018} and \citet{Martin2020}, who introduced the concept of graph networks for unsupervised morphological analysis. \citet[][hereafter C21]{Cheng2021a} investigated how galaxy morphologies can be categorised using a fully UML framework composed of a variational autoencoder structure. With UML approaches, C21 demonstrated that machine-based classifications can separate galaxies with different physical properties, such as galaxy colour, based solely on their structure using monochromatic images. This not only highlights the limitations of conventional visual classification in establishing clear decision boundaries for transitional morphological features, but also indicates that visual inspection may overlook subtle details in galaxy morphology which capture the footprint of galaxy evolution (see also \citealt{Cheng2021b}; for reviews on ML applications to galaxy studies and unsupervised techniques in astronomy, see \citealt{Huertas-Company2023,Lieu2024,Fotopoulou2024}).

    Here, we employ self-supervised ML techniques, such as an BYOL \citep{Grill2020,Walmsley2022} and autoencoder \citep{Rumelhart1986,Baldi2012,Bank2020}, to minimize human biases and identify extractable features from images using machine vision techniques. In this work, we focus on an autoencoder framework. Previous successful applications, such as \citet{Cheng2020b,Cheng2021a}, pioneered the use of an autoencoder framework to extract representative and interpretable galaxy structural features for unsupervised ML classifications. Since then, using an autoencoder framework for feature extraction has become a common approach for similarity search, anomaly detection, morphological categorisation and classification of galaxies in this field \citep[e.g.][etc]{Hayat2021,Zhou2022,Xu2023,Seo2023,Tohill2024}. Despite the growing popularity of self-supervised ML approaches, certain choices -- such as the selection of the number of latent features to use in the architectures and the interpretation of these features -- remain unclear. 
    
    We propose a combination of the autoencoder framework and principal component analysis (PCA) to recombine the features that capture most of the variance in the data. In addition to exploring and understanding the visual features directly extracted from images through machine vision, we will investigate whether an optimal number of features exists to capture meaningful representations, and how these can be linked to established structural measurements and physical properties of galaxies. 

    This paper is organised as follows. \S~\ref{sec:data} introduces the simulated datasets used and explains the data preparation steps used for model training. \S~\ref{sec:methodology} outlines the methodology, including the introduction of the autoencoder and PCA techniques. \S~\ref{sec:result_latent_features_VAE} first discusses the features learned through the VAE framework, followed by the results of applying PCA to the VAE-learned features in \S~\ref{sec:vae+pca}. \S~\ref{sec:subset_training} examines the extracted features for subsamples of the data with specific morphologies. In \S~\ref{sec:physical_comparison} we construct machine vision through the learned features and compare it with physical properties. Finally, \S~\ref{sec:conclusion} summarises our findings and outlines potential directions for future applications.

\section{Data Set}
\label{sec:data}
    
Here, we analyse images of simulated galaxies. The simulation and image construction are described in \S~2.1 and the preparation of these images for analysis is outlined in \S~2.2.

        \subsection{The EAGLE simulation}
        \label{sec:eagle}
        
        The Evolution and Assembly of GaLaxies and their Environments \citep[\texttt{EAGLE};][]{Schaye2015} is a suite of hydrodynamical cosmological simulations designed to model the formation and evolution of galaxies in a $\Lambda$-cold dark matter universe.
        The runs use the best-fitting cosmological parameters from \citet{Planck2014}. The simulations were run using a modified version of the \texttt{GADGET-3} code \citep{Springel2001,Springel2005}. %The modifications include an improved smoothed particle hydrodynamics (SPH) algorithm known as \textsc{anarchy}, which incorporates a pressure-entropy formulation of SPH \citep[][]{Hopkins2013}, enhanced treatments of viscosity and conduction \citep[][]{Cullen2010, Price2008}, and improved subgrid physics models to capture unresolved processes such as radiative cooling \citep[][]{Wiersma2009a}, reionisation \citep[][]{HaardtMadau2001}, star formation \citep[][]{SchayeDallaVecchia2008}, stellar mass loss and metal enrichment \citep[][]{Wiersma2009b}, energy feedback from star formation \citep[][]{Vecchia2012}, black holes \citep{Rosas-Guevara2015}, and active galactic nuclei (AGN) \citep[][]{Booth2009}. 
        The simulations rely on sub-grid models to capture physical processes that are only partially resolved and these models require parameters to be set. The parameter values  were calibrated so that the simulation output matched the observed $z\sim0$ galaxy stellar mass function \citep{Crain2015}, galaxy size distribution \citep[compared with observations,][]{Shen2003,Baldry2012}, and the scaling relations between supermassive black holes and stellar mass \citep{McConnell2013}. Full details of the \texttt{EAGLE} simulation can be found in \citet[][]{Schaye2015}.

        We use mock optical images created by \citet{Trayford2017}, sampled from the $z=0.1$ snapshot of the Ref-L0100N1504 simulation. The Ref-L0100N1504 simulation has the largest volume in the \texttt{EAGLE} suite with a box size of 100 comoving Mpc (cMpc), and contains the greatest number of galaxies. However, it also has the lowest resolution, with an initial baryonic particle mass $m_\mathrm{g} = 1.81 \times \,10^6 \,\mathrm{M}_{\odot}$, and a dark matter particle mass $m_\mathrm{dm} = 9.70 \times \,10^6 \, \mathrm{M}_{\odot}$. The gravitational softening is held fixed in comoving units from the initial conditions down to $z=2.8$ and is fixed in proper units thereafter. The Plummer-equivalent gravitational softening length in proper units is $\epsilon = 0.70\, \mathrm{kpc}$. This means that we should focus our attention on more massive galaxies, which have characteristic length scales that exceed the smoothing introduced by the gravitational softening. 
        
        The photometry of the mock observed images is produced by combining the \textsc{galaxev} stellar population synthesis models \citep{Bruzual2003} with the star formation and chemical enrichment histories predicted by \texttt{EAGLE}, following the implementation described in \citet{Trayford2015}. This yields Sloan Digital Sky Survey (SDSS) $ugriz$ band  photometry for each simulated galaxy. Dust attenuation is modelled using the \texttt{SKIRT} Monte Carlo 3D radiative transfer code \citep[][]{Baes2003, Baes2011,Camps2015}. In this code, the cool, metal-enriched gas serves as a tracer for dust in the diffuse interstellar medium.  \texttt{SKIRT} is computationally expensive and so is only run for the most massive galaxies. Integral field unit (IFU) data cubes are generated for dust-attenuated galaxies with $\mathrm{M}_* > 10^{10}\, \mathrm{M}_{\odot}$. Broad-band mock SDSS images are then produced by integrating over the spectral dimension of the data cubes using the $ugriz$ filters, and are also provided in three-color portable network graphics (PNG) format using $gri$ filters. Full details of the construction of the mock SDSS images and photometry can be found in \citet{Trayford2017}.  
        
        %Our analysis applies on only well-resolved galaxies with a stellar mass of $\mathrm{M}_* > 10^{10} \mathrm{M}_{\odot}$, which corresponds to $\gtrsim10^4$ stellar particles. 
        %and represent photometry in the SDSS $u$, $g$ and $r$ filters, simulated at a distance of 20Mpc.

        %To ensure only well-resolved galaxies are included in the production of mock images, a stellar mass cut of $\mathrm{M}_* > 10^{10} \mathrm{M}_{\odot}$ has been applied. 
        
        %The dust model applied in these images accounts for attenuation due to both diffuse interstellar dust and birth-cloud dust \citep{Charlot2000}. The generated mock images have a dimension of $256\times256$ pixels. 

        %The mock images have been created by \citep[][]{Trayford2017} and are formed using mock fluxes corresponding to the SDSS $u$, $g$ and $r$ band filters, at a distance of 20Mpc. To more accurately mimic observations, The SKIRT Monte Carlo radiative transfer code \citep[][]{Baes2003, Baes2011} is used to model dust by calculating a three-dimensional radiative transfer of each galaxy, and using the cool, enriched gas as a tracer of dust in the diffuse ISM. This results in 3-band images with a resolution of 256$\times$256.

        \subsection{Data preparation for model training}
        \label{sec:pre-processing}
            We analyse the three-colour PNG images, which mimic $gri$-band SDSS images. Each PNG image has a dimension of $256\times256$ pixels, and is only produced for dust-attenuated galaxies with $\mathrm{M}_* > 10^{10} \,\mathrm{M}_{\odot}$ due to the computational expense. To ensure reliable model performance and reduce sources of bias, several pre-processing steps are applied to the dataset prior to ML model training. These steps include: (1) pixel normalisation and (2) data augmentation. %(1) initial data selection, (2) pixel normalisation, and (3) data augmentation. %and dataset balancing through data augmentation.

            %\subsubsection{Initial data selection}
            %\label{sec:data_selection}
            %The initial selection is performed using flags obtained from S\'ersic profile fitting, as provided in \citet{deGraaff2021}. Note that these numbers are not the values of the S\'ersic index itself, but instead these flags indicate different conditions or issues encountered during the fitting process -- for example, a flag value of 1 indicates that the fit converged at a predefined boundary; a value of 2 denotes fits involving multiple components; and a value of 4 corresponds to poor fitting results, assessed through visual inspection. Detailed description of the flags can be checked in \citet{deGraaff2021}. To help better interpret the extracted features from galaxy images using ML approaches, this initial selection of the sample using S\'ersic fitting flags results in 41 galaxies being excluded from further analysis. The images rejected in this way generally contain overly complex and/or subtle structures that lead to unreliable parametric fits. Applying this initial selection to the sample results in 3\,583 galaxies remaining in our dataset.  
            %Several (41) of the mock images in the dataset are flagged for `bad' fit images during Sersic profile fitting by \citet[][]{deGraaff2021}. 
            %These flagged cases include fits which have converged at the allowed boundary of Sersic index, and those identified through visual inspection of the fits and residual images. All of these 'bad' fit images have been removed from the dataset.    

            \subsubsection{Pixel normalisation}
            \label{sec:data_normalisation}
            The initial PNG images consist of $gri$-band images, where the pixel values in each band are relative and represent flux differences (i.e., colours) between bands. Here, we aim to train ML models to identify and learn features that correlate with galaxy morphologies. Although there are known correlations between galaxy morphology and colour, the relationship is not unique -- galaxies with similar colours can exhibit diverse morphologies, and vice versa. Hence, to remove any contribution from the colour during ML model training, we therefore normalise each galaxy image independently in each band to fall in the range of 0 to 1. %each of the three bands in every image have been independently normalised to the range of 0 and 1. 
            This negates any potential bias from relative pixel values between bands and encourages the model to prioritise the underlying morphological structures and patterns within the images.

            \subsubsection{Galaxy structure and data augmentation}
            \label{sec:data_augmentation}
            %\textcolor{red}{
            Galaxies are often structurally decomposed into two distinct components: a rotationally supported disk and a pressure-supported bulge. Galaxy structures change across the spectrum between disk and bulge-dominated morphologies. Here, we use the disk-to-total stellar mass ratio ($D/T$), which is derived from the mass fraction of rotationally supported stars (i.e. those belonging to the disk component), as a kinematic diagnostic to distinguish between disk and bulge-dominated structures \citep{Thob2019}. This quantity can be estimated in a simulation, where full velocity information is available. Moreover, as this quantity is an intrinsic property of galaxies that do not differ by image quality, it better reflects the true underlying structures of galaxies. 
            
            A galaxy in which a significant portion of the stellar mass resides in the disk, with $D/T>0.2$, is classified as disk-dominated. This category generally includes spiral galaxies. Galaxies with a negligible disk component, characterised by $D/T<0.1$, are labelled as bulge-dominated and include spheroidal and elliptical galaxies. Those with intermediate values, i.e. $0.1\leq{D/T}\leq0.2$, have transitional morphologies, blending the features of both disks and bulges. These three structural types represent the fundamental morphological classes of galaxies and can often be visually identified in imaging data. The initial data set contains 3\,548 galaxies, excluding the small number with negative $D/T$ values, which appear due to the calculation method used in \citet{Thob2019}. Of these mock images, 2\,906 ($\sim$82\%) have disk-dominated structures with $D/T>0.2$, while 323 ($\sim$9\%) galaxies are bulge-dominated with $D/T<0.1$. The remaining 319 ($\sim$9\%) galaxies have intermediate structures with $0.1\leq{D/T}\leq0.2$, reflecting ambiguous morphological features between typical the disk and bulge-dominated regimes.
            
            Note that the classification here is intended to mitigate potential biases arising from an uneven number of samples representing different structures during model training. For instance, the significantly larger number of disk galaxies in the sample can result in biases during the feature learning process, as the losses become dominated by the reconstruction of galaxy images with disk structures and the features that characterise them. Such class imbalance issues are known to hinder the generalisability of learned representations, especially for underrepresented morphological types \citep[see the discussion on imbalanced data in][]{Cheng2020a}. Therefore, a morphological definition based on intrinsic dynamical information provides a more robust foundation to balance the number of samples with intrinsically different structures. To resolve imbalance, we employ data augmentation to increase the number of galaxies with the under-sampled structures, i.e. bulge/elliptical and intermediate structures in this case. This enables more robust and comprehensive feature learning across the spectrum of disk and bulge components. 

            The data augmentation process rotates the original mock images using the \textsc{Keras} library function \texttt{ImageDataGenerator}. 
            %Specifically, we generate new images from the original \textcolor{red}{'transitional'} and elliptical structure images, through processes such as rotating and flipping the original images. 
            %These transformations preserve the intrinsic galaxy structures while increasing sample diversity. 
            For bulge-dominated (elliptical) galaxies, each original image is used to generate 8 new images, resulting in a total of 2\,907 images with this type ($323 \times 8 + 323$). Similarly, each of the 319 galaxies with intermediate structures is used to generate 8 new images, yielding a total of 2\,871 images representing transitional morphologies ($319 \times 8 + 319$). The final augmented datasets hence consist of 8\,684 images with approximately equal fractions of disk/spiral ($\sim$33.5\%), bulge/elliptical ($\sim$33.5\%), and intermediate ($\sim$33.0\%) structures.

\section{Methodology}
\label{sec:methodology}
    
    We use self-supervised ML in the form of a variational autoencoder \citep[VAE;][\S~\ref{sec:vae}]{Kingma2014} to reduce the dimensionality of galaxy images into a smaller number of representations. %By using a convolutional architecture, the 
    The VAE effectively learns a set of latent features\footnote{Throughout, we use the term `latent features' to refer to the representations learned by the VAE, while the term `extracted features' is used for those obtained after the PCA process.} %The features are called latent as they are not direclty observable in an image, like a spiral arm, but instead are revealed after performing an analysis on the image.} %(i.e. features that are not explicit or are hidden) 
    that compactly represent the key characteristics of a galaxy from imaging data, and the original image can then be reconstructed using these features. %, with the accuracy of the reproduction quantified using a metric. 
    The latent features obtained from the VAE will be further distilled using principal component analysis \citep[PCA;][\S~\ref{sec:pca}]{Wold1987}, recombining the latent features to form a new set of extracted features that are ordered by the variance they account for in the images. %This enhances interpretability and helps us to understand how the features are connected with the known astronomical properties of galaxies.
    
    \subsection{Variational autoencoder}
    \label{sec:vae}
       
        An autoencoder is an artificial neural network and is essentially the combination of two models, an encoder and a decoder. Generally speaking, the encoder takes an image, and breaks it into a set of discrete latent features, and as such, each image can effectively be described by a set of latent features. The decoder does the opposite; it attempts to reconstruct the original image using the learned features derived from the encoder. The performance of the image reconstruction process is assessed using an accuracy measure, such as the mean squared error (MSE) or binary cross-entropy (BCE),  between the input and reconstructed pixel values. %, derived from the reconstruction process. 
        The latter statistic is more suitable for normalised image pixel values, as it models each pixel as a Bernoulli-distributed variable and provides pixel-wise probability estimates. %(more details in \S~\ref{sec:vae_formulas}). 
        We rescale the pixel values of each band to be in the range 0 to 1 (see \S~\ref{sec:data_normalisation}). Therefore, we choose BCE as the cost function to guide the model optimisation during model training. 

        The advantage of a VAE is that it incorporates a probabilistic approach to the encoding process, instead of producing a set of discrete latent features like an autoencoder. The latent features of a VAE are typically sampled as Gaussian distributions, parameterised by their mean and variance. %, which are then stochastically sampled to the latent features. 
        %This provides more structured latent features, facilitating better interpretability. %Additionally, a structured posterior gives another advantage as a generative model -- one can draw a random set of values from this Gaussian-like posterior to generate a new image. 
        The choice of a Gaussian prior offers a simple, yet effective approach to regularising the latent space, resulting in more structured latent features and improved interpretability. %More details of the formulation are introduced in \S~\ref{sec:vae_formulas}. %Also, by sampling the resulting Gaussians, it is straightforward to generate new images with the same properties as the original set, allowing for easy data augmentation. %allowing us to more closely explore the visual representation of the extracted features.

        \subsubsection{Key formulae}
        \label{sec:vae_formulas}
        
            The training of a VAE model is guided not only by the reconstruction accuracy but also by forcing each latent feature to follow the distribution of a prior. The total VAE loss is then quantified as the sum of two terms, the reconstruction loss and a regularisation term, which takes the form of the Kullback-Leibler (KL) divergence, $D_\mathrm{KL}$ (explained below):   
            \begin{equation}
                \label{eq:vae_loss}
                L_\mathrm{VAE} = L_\mathrm{recons} + \beta D_\mathrm{KL}. 
            \end{equation}
            The reconstruction loss, $L_\mathrm{recons}$, ensures that the decoded output resembles the input image, while $D_\mathrm{KL}$ regularises the learned distributions of latent features (i.e. posterior distributions). %contains two terms: (1) the Kullback–Leibler (KL) divergence \citep[Equation~\ref{eq:kl};][]{Kullback1951}, $D_{\mathrm{KL, Gaussian}}$,  which regularises the distributions of latent features (posterior distributions) to match with the prior Gaussian distribution; and (2). 
            The $\beta$ value is used to balance the contributions of the two loss terms\footnote{This form of VAE is called `$\beta-$VAE' \citep{Higgins2016, Burgess2018}.}. We set the $\beta$ value to 0.0001, which was selected jointly with the number of latent features used in the VAE through a grid search to achieve a balance between the two loss terms.
            
            For the reconstruction loss, as mentioned in \S~\ref{sec:vae}, we implement the BCE loss function as follows, instead of MSE:
            \begin{equation}
                \label{eq:reconstruction_loss}
                L_{\mathrm{recons}} = -\frac{1}{N_{\mathrm{p}}}\sum_{i=1}^{N_{\mathrm{p}}} \bigl( y_{i}\ln{\hat{y}_{i}} + (1 - y_{i})\ln{(1 - \hat{y}_{i})} \bigr),
            \end{equation}
            where $N_{\mathrm{p}}$ represents the total number of pixels, $y_i$ denotes the original pixel value at the $i-$th pixel, while $\hat{y_i}$ is the predicted pixel value at the $i-$th pixel, also interpreted as a probability. Here, the pixel values of the input images are normalised to the range 0 to 1 to eliminate any bias caused by varying intensity values (\S~\ref{sec:data_augmentation}), and the output layer uses a sigmoid activation function (\S~\ref{sec:vae_architecture}), which generates probabilistic output pixel values between 0 and 1. The choice of BCE is advantageous for this setup, as it treats each pixel as a Bernoulli-distributed variable, which better represents the nature of the normalized pixel values, enabling more accurate reconstruction compared to MSE, which assumes a Gaussian distribution. 
            
            On the other hand, for $D_{\mathrm{KL}}$, the VAE employs the KL divergence \citep[Equation~\ref{eq:kl};][]{Kullback1951} to regularise the latent feature distributions, encouraging them to approximate a prior of a standard Gaussian distribution with mean value of 0 and standard deviation of 1. 
            %, $D_{\mathrm{KL, Gaussian}}$,  which regularises the distributions of latent features (posterior distributions) to match with a prior Gaussian distribution.   
            %The KL divergence quantifies the similarity between the latent feature (posterior) distributions and the prior distributions. The typical VAE uses a standard Gaussian distribution with the mean value of 0 and standard deviation of 1 for the prior. 
            This is calculated as follows:
            \begin{equation}
               \label{eq:kl}
                D_{\mathrm{KL}} = \frac{1}{2N_{\mathrm{f}}} \sum_{{\mathrm{j}}=1}^{N_{\mathrm{f}}}{} \bigl( \mu_{\mathrm{j}}^2 + \sigma_{\mathrm{j}}^2 - \ln{\sigma_{\mathrm{j}}^2} - 1 \bigr),
            \end{equation}
            where $N_{\mathrm{f}}$ represents the total number of latent features. For the ${\mathrm{j}}-$th latent feature, a Gaussian distribution with a mean $\mu_{\mathrm{j}}$ and a standard deviation $\sigma_{\mathrm{j}}$ is used to sample the latent feature. %Note that the Equation~\ref{eq:kl} is a simplified form, derived under the assumption that the target distribution is a standard Gaussian distribution with the mean value of 0 and standard deviation of 1. 

        \subsubsection{Model architecture}
        \label{sec:vae_architecture}
            \begin{figure*}
            %\centering
            \includegraphics[width=0.9\textwidth]{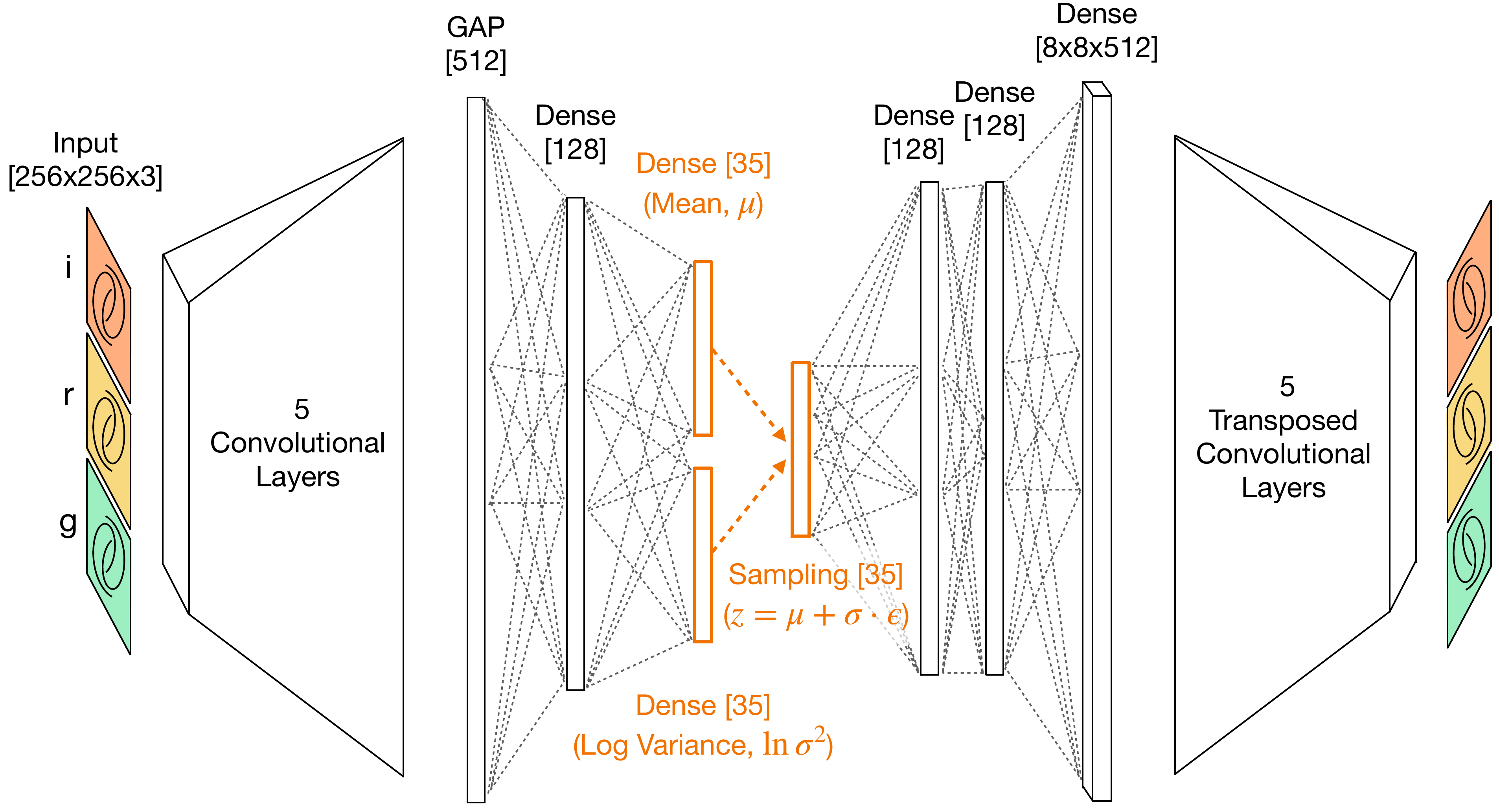}
            \caption{A schematic architecture of the VAE used in the feature learning process. The orange blocks in the intermediate layers represent the latent features, sampled using the means and log-variances.}
            \label{fig:vae_architecture}
            \end{figure*}
            
            \begin{figure*}
            %\centering
            \includegraphics[width=0.9\textwidth]{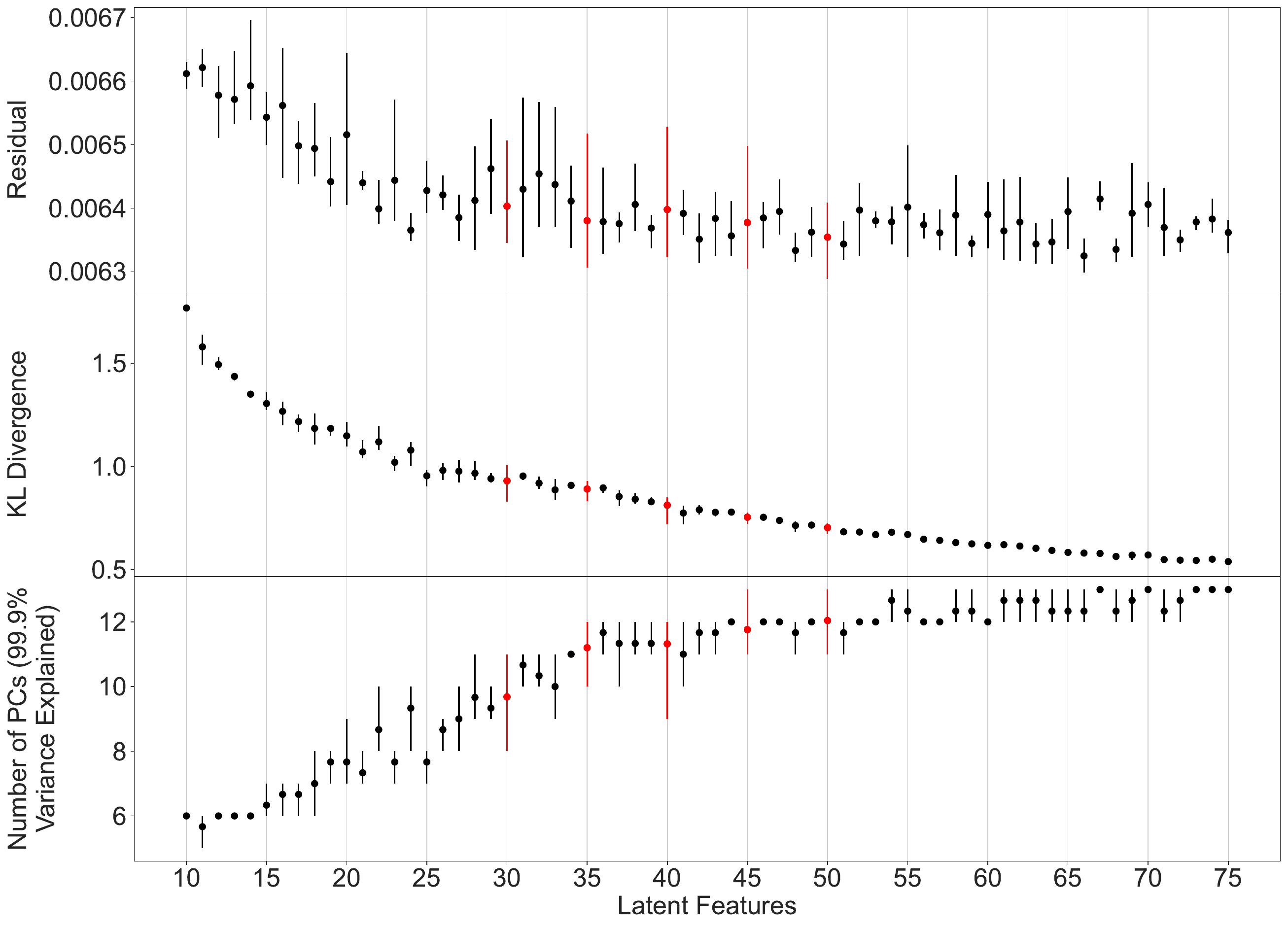}
            \caption{The performance of the VAE as a function of the number of latent features retained. The metrics are the residual between the input and reconstructed images (top), the KL divergence (middle), and the number of principal components (PCs) explaining 99.9 per cent of the variance in the data (bottom). Each black dot represents the mean values of three runs for a specific number of latent features, while red dots correspond to cases with 25 runs. In each run the initial weights for the neural networks are chosen randomly. Error bars indicate the uncertainty, defined by the minimum and maximum values of the metric calculated across multiple runs.}
            \label{fig:loss_plots}
            \end{figure*}

            The architecture of the VAE is shown in Fig.~\ref{fig:vae_architecture}, and the details of each layer are given in Table~\ref{autoencoder_architecture}. In the encoder, five convolutional layers are used to gradually reduce the spatial size of the input images from 256$\times$256 pixels to 8$\times$8 pixels, while progressively increasing the number of filters from 3 to 512. This setup helps stabilise the chosen parameters during the transformation process, ensuring smoother learning and more consistent feature extraction. A global average pooling (GAP) layer follows to further reduce the dimensionality of the feature map from 8$\times$8$\times$512 to a one-dimensional latent representation with a size of 512 parameters. This is carried out by averaging the parameter values of each 8$\times$8 filter from the previous layer. This reduction in the number of parameters helps prevent overfitting during the training process. Then, dense layers are used to connect each parameter. The number of neurons in the dense layers gradually decreases from 512 to 35, which represents the number of latent features obtained by the VAE. This number is determined jointly with the value of $\beta$ in Eqn~\ref{eq:vae_loss} through a grid search by observing the point at which the residual between input and reconstructed images reaches a plateau, while the corresponding KL divergence term is approximately unity (see Fig.~\ref{fig:loss_plots}). A KL divergence value of 1 indicates that the latent distributions retain information beyond the prior while remaining sufficiently close to the prior. This criterion ensures a balance between effective representation learning alongside the regularisation of the distribution modelling. We note that generating additional latent features is of little  benefit as it reduces interpretability, as the distributions begin to duplicate the prior. Additionally, the bottom panel of Fig.~\ref{fig:loss_plots} shows the relationship between the number of latent features and the number of principal components (PCs) required to explain 99.9 per cent of the variance in the data. This demonstrates that using more than 35 latent features is redundant, as any additional features are subsequently discarded by the PCA process (see \S~\ref{sec:pca} for an introduction of the PCA technique and \S~\ref{sec:vae+pca} for the discussion of results). Moreover, this highlights a specific range on the number of extracted features actually required to explain 99.9 per cent of the variance in the data.

            The latent vectors are composed of two components -- the mean ($\mu$) and log-variance ($\ln\sigma^{2}$), defining the initial posterior as a Gaussian distribution. As mentioned in Section~\ref{sec:vae_formulas}, these latent features are expected not only to represent the input images well for high-fidelity reconstruction but are also regularised to approximate a Gaussian distribution. The former condition is met by the reconstruction loss (Equation~\ref{eq:reconstruction_loss}) and the latter is realised by the KL divergence (Equation~\ref{eq:kl}). These Gaussian distributions are stochastically sampled (in the sampling layer) to form a single latent vector, $z$. This is done through the reparameterisation trick, $z = \mu + \sigma\cdot\epsilon$ with random variable  $\epsilon \sim \mathcal{N}(0,1)$. This ensures that the cost function remains differentiable while introducing stochasticity, allowing the VAE to learn and generate diverse variations of the data. %\textcolor{red}{The latent vector is then transformed through 3 layers of planar flows (as described in Section \ref{sec:vae_formulas}) to allow more expressive latent distributions.} 
            
            The decoder is essentially the inverse of the encoder, except that the GAP layer does not have a direct inverse. Instead, we use dense layers to bring up the dimensionality, %to 32\,768 
            which can then be transformed into the dimensionality of 8$\times$8$\times$512, to match the output of the final convolutional layer in the encoder. Transposed convolutional layers are then used to upsample the latent representations, transforming the dimensionality back to 256$\times$256$\times$3, matching the size of the original input images.

            All convolutional and transposed convolutional layers have a kernel size of 3$\times$3. %to capture fine image details. 
            Both (transposed) convolutional layers and dense layers employ the Rectified Linear Unit \citep[\texttt{ReLU};][]{Nair2010_cs} activation function, such that $f\left( x \right)=0$ if $x<0$ while $f\left( x \right)=x$ if $x\geq0$, except for the final transposed convolution layer. This layer instead uses the sigmoid activation function: $f\left( x \right)=1/\left(1 + e^{-x}\right)$, to generate probabilistic outputs in the range of 0 and 1, compatible with its use in the BCE loss.

            \begin{table}
            \centering
            \begin{tabular}{c c c}
                    Layer Type & Spatial Size & Filters \\
            
                    \hline\hline
    
                    Input Layer & 256$\times$256 & 3 \\
                    Convolutional & 128$\times$128 & 32 \\
                    Convolutional & 64$\times$64 & 64 \\
                    Convolutional & 32$\times$32 & 128 \\
                    Convolutional & 16$\times$16 & 256 \\
                    Convolutional & 8$\times$8 & 512 \\
                    Global Average Pooling & 512 \\
                    Dense & 128 \\
                    Dense (Mean) & 35 \\
                    Dense (Log Variance) & 35 \\                  
                    Sampling & 35 \\
                    %Planar Flow & 30 \\
                    %Planar Flow & 30 \\
                    %Planar Flow & 30 \\
            
                    \hline
    
                    Input Layer & 35 \\
                    Dense & 128 \\
                    Dense & 128 \\
                    Dense & 512 \\
                    Dense & 32768 \\
                    Reshape & 8$\times$8 & 512 \\
                    Transposed Convolutional & 16$\times$16 & 256 \\
                    Transposed Convolutional & 32$\times$32 & 128 \\
                    Transposed Convolutional & 64$\times$64 & 64 \\
                    Transposed Convolutional & 128$\times$128 & 32 \\
                    Transposed Convolutional & 256$\times$256 & 3 \\

                \end{tabular}
                \caption{Summary of each of the layers of the autoencoder including the type of layer and the dimensionality of the output, comprised of spatial size and number of filters. The top section represents the encoder and the bottom section represents the decoder.}
                \label{autoencoder_architecture}
            \end{table}

    \subsection{Principal component analysis}
    \label{sec:pca}

        Principal component analysis (PCA) is a statistical technique used to reduce the dimensionality of data while preserving as much variance as possible \citep{PCA1990}. It transforms the original features, $z$, onto a new set of orthogonal axes, called `principal components', which are ordered by the amount of variance they capture. We apply PCA through singular value decomposition (SVD) of the mean-centred latent feature matrix $\mathbf{Z}^{\prime} \in \mathbb{R}^{N_{\mathrm{s}} \times N_{\mathrm{f}}}$, where $N_{\mathrm{s}}$ is the number of samples and $N_{\mathrm{f}}$ is the number of latent features. Each row of $\mathbf{Z}^{\prime}$ corresponds to a mean-centred latent feature vector, with $z_{\mathrm{kj}}^{\prime}=z_{\mathrm{kj}}-\bar{z_\mathrm{j}}$, where $z_{\mathrm{kj}}$ represents the $j-$th feature of the $k-$th sample and $\bar{z_{\mathrm{j}}}$ denotes the mean value of the $j-$th features across all samples. The SVD of $\mathbf{Z}^{\prime}$ is given by: 
        \begin{equation}
        \label{eq:svd}
            \mathbf{Z}^{\prime} = \mathbf{U} \mathbf{\Sigma} \mathbf{V}^{\top},
        \end{equation}
        \noindent where $\mathbf{U} \in \mathbb{R}^{N_{\mathrm{s}} \times N_{\mathrm{s}}}$ and $\mathbf{V} \in \mathbb{R}^{N_{\mathrm{f}} \times N_\mathrm{f}}$ are orthogonal matrices, and $\mathbf{\Sigma} \in \mathbb{R}^{N_{\mathrm{s}} \times N_{\mathrm{f}}}$ is a diagonal rectangular matrix containing the singular values. %$s_{m}$, for $m=1,2,3,...,N_f$ given that $N_f<N_s$. 
        $\mathbf{V}$ contains the principal components (PCs) -- the orthogonal directions of maximum variance in the feature space -- obtained as the eigenvectors of $\mathbf{Z}^{\prime \top} \mathbf{Z}^{\prime}$ (the unnormalised convariance matrix of $\mathbf{Z}$). Note that the normalised convariance matrix, $C_\mathbf{Z}$, is scaled as $C_\mathbf{Z}=\mathbf{Z}^{\prime \top} \mathbf{Z}^{\prime}/\left(N_{\mathrm{s}} - 1\right)$. The squared roots of the corresponding eigenvalues contribute to the singular values in the diagonal rectangular matrix $\mathbf{\Sigma}$, quantifying the variance along the PC axes. The $\mathbf{U}$ can then be computed with the obtained $\mathbf{Z}^{\prime}$, $\mathbf{V}$, and $\mathbf{\Sigma}$. The columns of the matrix $\mathbf{U}$ are the eigenvectors of $\mathbf{Z}^{\prime} \mathbf{Z}^{\prime \top}$, and each row of the product $\mathbf{U}\mathbf{\Sigma}$ provides the coordinates of each sample in the PC axes. %, describing how each sample is projected onto the PC basis. 
        Namely, the new latent features on the PC basis, $\mathbf{Z_{\mathrm{{PCA}}}}$, is given by
        \begin{equation}
            \nonumber
            \mathbf{Z_{\mathrm{{PCA}}}}=\mathbf{U}\mathbf{\Sigma}=\mathbf{Z}^{\prime}\mathbf{V}.
        \end{equation}
        \noindent To reduce dimensionality, we retain the top $m$ singular values and their corresponding principal components, $\mathbf{V}_m \in \mathbb{R}^{N_s \times m}$, which contribute to the majority of the variance. The selected features, $\mathbf{Z}_{\mathrm{PCA},m}$, can then be computed by $\mathbf{Z}^{\prime}\mathbf{V_m}$. Here, we provide a mathematical description of the PCA computation. The selection of the optimal number of PCs, i.e. extracted features, that capture a sufficient amount of variance to effectively represent the visual appearance of galaxies in images is discussed in \S~\ref{sec:optimal_number_of_extracted_features}.

    \section{Latent Features Learned by VAE}
    \label{sec:result_latent_features_VAE}
    \begin{figure*}
            %\centering
            \includegraphics[width=\textwidth]{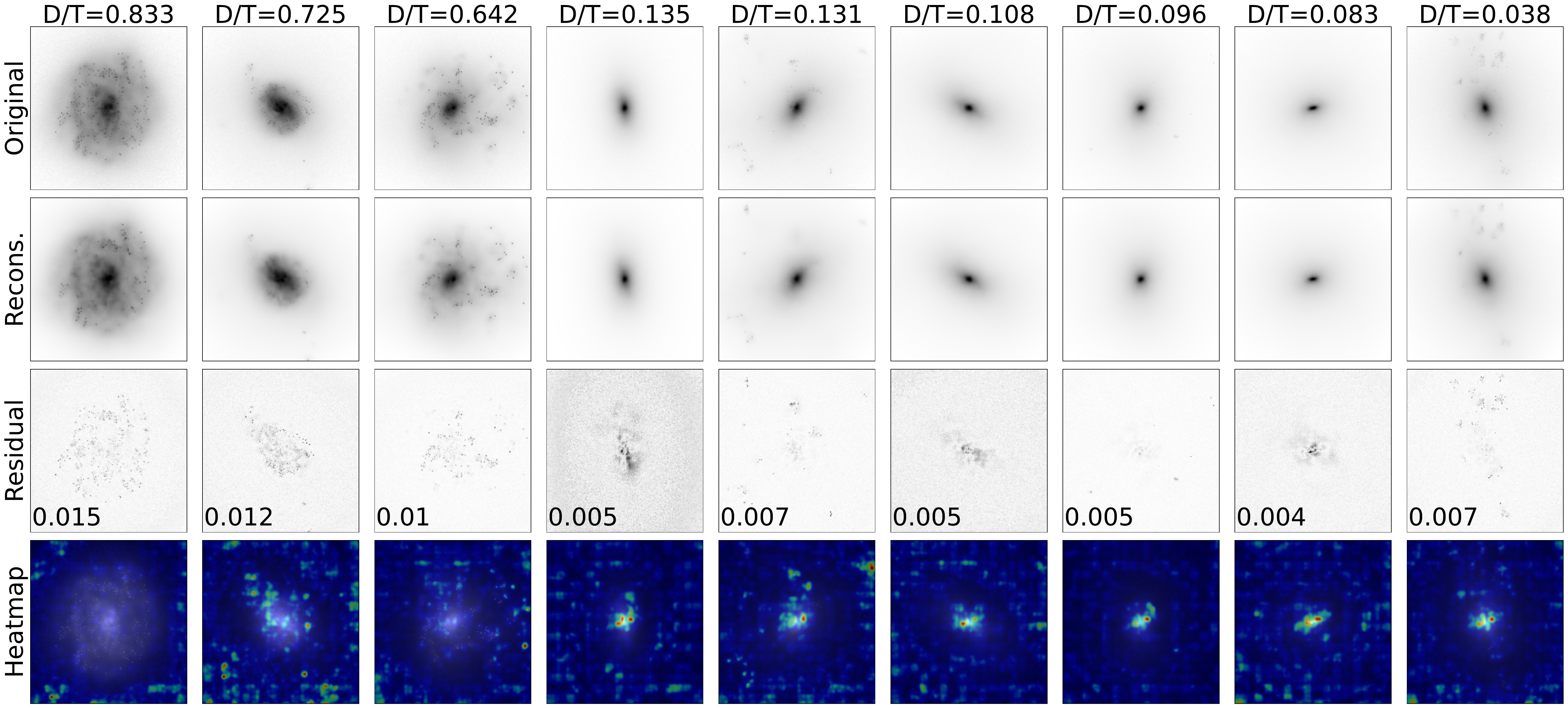}
            \caption{Examples of galaxy images grouped by morphological type: from left to right disk-dominated ($D/T>0.2$), intermediate ($0.1\leq{D/T}\leq 0.2$), and bulge-dominated ($D/T<0.1$). Above each panel, the $D/T$ value is provided for each galaxy. The top and second rows show the original and reconstructed images, respectively. The third row presents the residuals between the original and reconstructed images, and the residual values are provided at the corner. The bottom row displays feature weight maps, highlighting the regions that contribute most to the reconstruction with the reddest  colour shading.}
            \label{fig:vae_example}
    \end{figure*}
    \begin{figure*}
            %\centering
            \includegraphics[width=\textwidth]{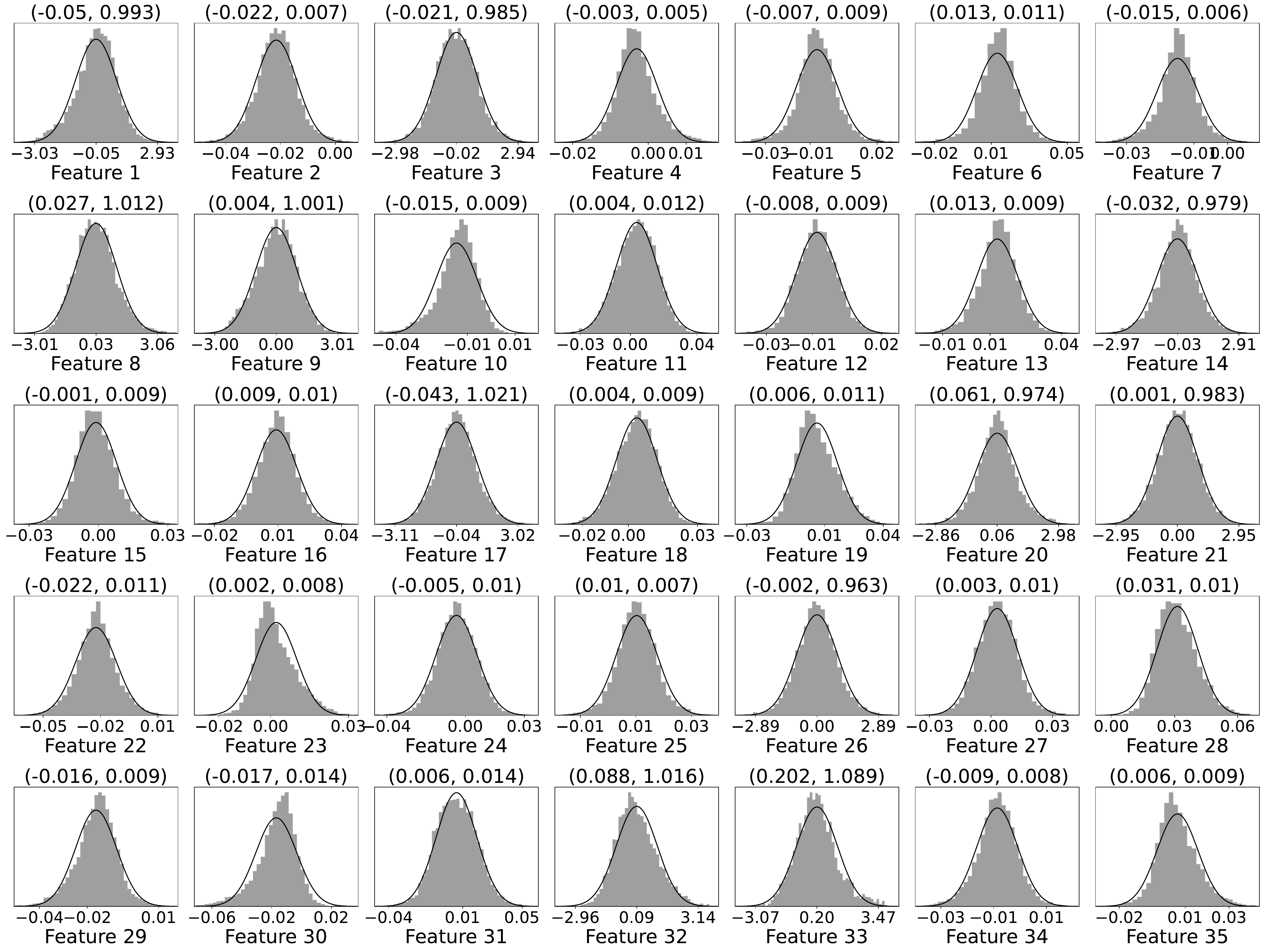}
            \caption{Distributions of the mean value ($\mu$) in the 35 latent feature vectors extracted by the VAE shown by the histograms. Above each panel, the mean (left) and standard deviation (right) of the corresponding feature's mean ($\mu$) values are shown in the parentheses. The solid black line in each panel shows the  Gaussian distribution plotted using the calculated mean and standard deviation values.}
            \label{fig:latent_distribution}
    \end{figure*}
    \begin{figure*}
            %\centering
            \includegraphics[width=\textwidth]{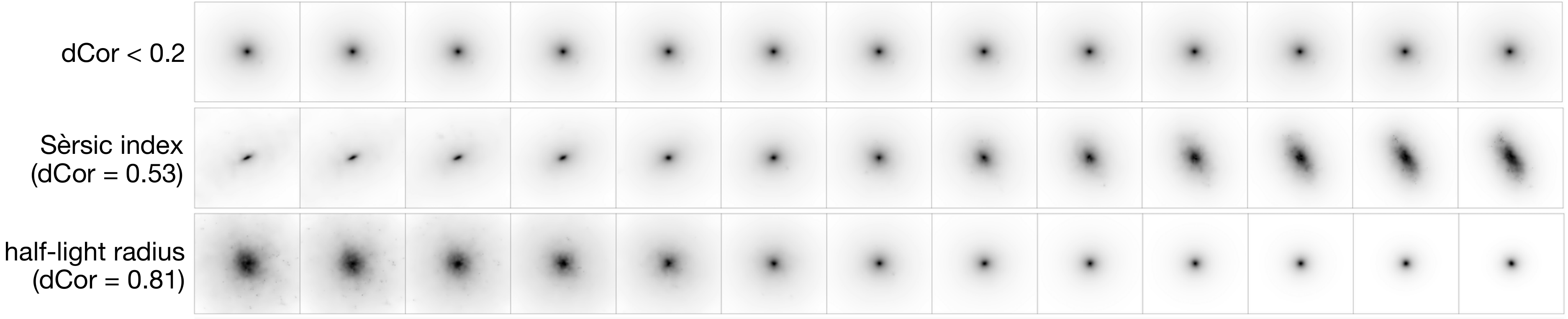}
            \caption{Example illustration of galaxy appearance transitions resulting from variations in individual latent features. In each row, a single latent feature is varied between the maximum and minimum values of this feature across the dataset, while all others are fixed at their mean values of all samples. The top row shows a feature with low correlation ($\mathrm{dCor}<0.2$) to any structural measurement. The second and third rows present features with the strongest correlations to S\'ersic index and half-light radius, respectively.}
            \label{fig:transition_somefeatures}
    \end{figure*}
    
    By design the VAE process yields 35 latent features that not only can be used to effectively reconstruct the dataset to within some acceptable loss but are also structured to map onto a standard Gaussian distribution (\S~\ref{sec:vae_architecture}). Reconstruction examples are shown in Fig.~\ref{fig:vae_example}. From left to right, we order the examples by their $D/T$ values, first showing disk-dominated galaxies ($D/T>0.2$), then intermediate ($0.1\leq{D/T}\leq0.2$), and finally bulge-dominated ($D/T<0.1$) structures. The original and reconstructed images are presented in the top and second rows, respectively, with the third row displaying the residuals between them. The residual value, shown in each panel, is computed as the mean of the difference of the pixel values between the original and reconstructed images. The bottom row presents feature weight maps %\footnote{\textcolor{red}{Feature weight maps are generated using a gradient-based attribution method, where the gradients of each latent feature with respect to the input pixels are computed to identify the regions that most strongly influence the VAE reconstruction.}} 
    \citep{Simonyan2013}, which highlight the regions of the images that contribute most significantly to the reconstruction process. We performed a total of 25 runs with 35 latent features, and the resulting feature weight maps consistently show that the VAE focuses on different image regions for different morphological types. Attention shifts from the outskirts in disky galaxies (left) to the inner/central regions in more bulge-dominated systems (right). Additionally, we notice enhanced attention along the spiral arm structures for some spiral galaxies.
    
    Regarding the distributions of latent features, while the features are assumed to follow a Gaussian prior, they do not replicate a standard normal distribution exactly. As described in \S~\ref{sec:vae_architecture}, the number of latent features in the VAE is determined by balancing the reconstruction residual and KL divergence terms. The dimensionality is chosen to have a corresponding KL divergence value of approximately one, indicating that appropriate deviations between the latent feature distribution and the prior are permitted and expected. One may question whether the intrinsic distributions of underlying galaxy characteristics, as imprinted in images, genuinely follow a Gaussian form. We also experimented with more flexible latent feature distributions using normalising flows \citep{Rezende2015,Kobyzev2020}. However, the resulting features remained predominantly Gaussian, and the implementation of normalising flows yielded only negligible improvements in reconstruction performance. To preserve interpretability of the learned features, we therefore opted to proceed with the standard VAE model. The distributions of each VAE latent feature are shown in Fig. ~\ref{fig:latent_distribution}. 

    In detail, Fig.~\ref{fig:latent_distribution} presents the distributions of the mean values ($\mu$) in latent feature vector $z$, where $z = \mu + \sigma\cdot\epsilon$ (see \S~\ref{sec:vae_architecture}). For each latent feature, we compute the empirical mean and standard deviation across the dataset, and construct a corresponding Gaussian distribution to assess whether the VAE-derived mean values approximately follow a Gaussian form. %Note that we neglect the influence of the VAE-derived variance ($\sigma^2$) on the distribution of each mean value ($\mu$), as it should not significantly affect the overall shape of the distribution. 
    Overall, the latent features follow a Gaussian shape with minor deviations. However, some latent features span a narrow range of values with small standard deviations of $\lessapprox0.01$. These features may represent `redundant features', contributing to reconstruction quality by capturing background noise, but providing limited information about galaxy structures and underlying astronomical properties. Nevertheless, we note that although many of these features contribute negligibly to the total variance and are discarded by the subsequent PCA, having a sufficient number of latent features is essential to preserve the ability of the VAE to reconstruct important visual features. As demonstrated in Fig.~\ref{fig:loss_plots}, the PCA analysis indicates that 10-12 features can capture 99.9 per cent of the variance, but at least 35 VAE latent features are required for the PCA process to extract them effectively. 

    \subsection{Correlation with structural measurements}
    \label{sec:vae_structures}
    
    We further compare the latent features with the structural measurements for this set of simulated galaxies as reported in \citet[][dynamical information: $D/T$]{Thob2019}, \citet[][using \texttt{GALFIT}: S\'ersic index, axis ratio, position angle]{deGraaff2022}, and \citet[][using \texttt{statmorph\footnote{\url{https://github.com/vrodgom/statmorph}}}: half-light radius, concentration, asymmetry, smoothness]{Bignone2020}. The structural measurements in the latter two studies are based on the mock images produced by \citet{Trayford2017}; \cite{Thob2019} use the particle data from the simulation in their analysis. 
    
    To evaluate the correlation between latent features and the aforementioned structural measurements, we use the distance correlation \citep[dCor;][]{Gabor2007}, calculated using the \textsc{Python} package  \texttt{dcor}\footnote{\url{https://github.com/vnmabus/dcor}} \citep[][]{software_2022_dcor, Ramos-Carreno2023_dcor}. Similarly, to the Pearson correlation coefficient \citep{LeeRodgers1988}, this measure quantifies the strength of the association between variables; however, unlike the case of the Pearson statistic, the distance correlation is not limited to detecting linear dependencies and can capture more general non-linear relationships. This metric is defined based on the distance covariance between the variables and is normalised by their respective distance variances. Therefore, it provides a value between 0 and 1, where 0 indicates statistical independence between the variables and 1 implies perfect dependence. Here, since the relationship between latent features and structural measurements may be non-linear, the use of the distance correlation is more appropriate for this evaluation\footnote{We also evaluated the relationships using the Pearson correlation coefficient. The results were broadly consistent with those from the distance correlation, but slightly weaker for some features.}. %We adopt an empirical threshold of \textcolor{red}{$\mathrm{dCor}=0.3$} here to define weak and strong correlations. This choice is inspired by the commonly used threshold in Pearson correlation for identifying moderate to strong linear relationships, as well as visual inspection of features' distributions and their contribution to the reconstructed images. 
    
    In the top row of Fig.~\ref{fig:transition_somefeatures}, we show the impact of a latent feature with a low distance correlation ($\mathrm{dCor}<0.2$) to any structural measurements on the visual appearance of galaxies. We find that these features which do not lead to noticeable changes in galaxy morphology are also the same `redundant features' characterised by tiny variances in their distributions, as discussed previously in connection with  Fig.~\ref{fig:latent_distribution}. Although some of these redundant features may exhibit stronger correlations with specific structural measurements, their impact on the image reconstruction is negligible, as they contribute only a minimal fraction of the total variance.
    
    %\textcolor{red}{If using a threshold of 0.2 to define weak and strong correlation, xx latent features exhibit weak correlations with all examined structural measurements, while xx, xx, xx, xx, xx, xx, and xx latent features show stronger correlations with S\'ersic index, axis ratio, position angle, half-light radius, concentration, asymmetry, and smoothness, respectively.} 
    More importantly, we find that a structural measurement can be associated with multiple latent features, and a latent feature may, in turn, be linked to several structural measurements. For example, approximately $14.5\pm1.0$ and $6.0\pm1.5$ latent features have strong distance correlations ($\mathrm{dCor}\gtrapprox0.3$) with the half-light radius (size) and the S\'ersic index, respectively. Meanwhile, some latent features are simultaneously associated with multiple structural measurements like the half-light radius, concentration, asymmetry, and smoothness of galaxies. This reflects a degree of degeneracy in the comparison between the two sets of properties (latent features and structural properties), which is expected for two reasons: (1) many structural properties are intrinsically correlated with one another; and (2) the VAE framework does not attempt any disentanglement between the latent dimensions, allowing multiple features to capture overlapping information. Moreover, it is important to emphasize that the VAE latent features and the aforementioned structural measurements represent fundamentally different quantities. Structural measurements are independently derived from galaxy images using predefined analytical formulae; they are designed to quantify a specific structural property. In contrast, the VAE latent features collectively encode the entire image content, with every feature contributing to a portion of the reconstruction. For instance, suppose feature 1 and feature 2 primarily control the generation of disk-like and bulge-like structures, respectively. A spiral galaxy would be expected to show a strong contribution from feature 1 and a weaker contribution from feature 2, whereas an elliptical galaxy would likely exhibit a negligible value for feature 1 and a more dominant contribution from feature 2. In this scenario, both features 1 and 2 would, by definition, show correlations with the Sérsic index. Thus, the VAE latent features are in fact not directly comparable to these structural measurements; instead, they only provide a complementary, data-driven representation that constructs the visual appearance, which in turn correlates with these structural quantities.
    
    The comparison here simply serves to provide insight into the types of structures that each latent feature may be capturing. In Fig.~\ref{fig:transition_somefeatures}, we illustrate what the latent features may capture -- the second and bottom rows of Fig.~\ref{fig:transition_somefeatures} show the impact of the latent features with strongest correlations to S\'ersic index and half-light radius, respectively, on the visual appearance of galaxies. Nevertheless, given the strong intercorrelations among latent features, a method for recombining them is required to enhance interpretability.
    
    %In Fig.~\ref{fig:}, we compare the latent features with the structural measurements reported in \citet{Thob2019}, \citet{Bignone2020} and \citet{deGraaff2022}. Except for Thob's work, the structural measurements in the latter two studies are based on mock images produced by \citet{Trayford2017}. 
    
    %\begin{itemize}
    %    \item FIGURES for this section:
    %        \item reconstruction: original, reconstruction, residual for three galaxies (disk, bulge, middle). 
    %        \item heated maps on a randomly selected (1) disk, (2) bulge, (3) transitional galaxies
    %        \item distributions of 30 features compared with a standard Gaussian (not sure yet)
    %        \item correlation coefficient plot with structural measurements (D/T, Sersic, size, position angle, axis ratio, CAS).
    %        \item transition figures of some features 
    %\end{itemize}

    \section{Feature transformation with Principal Component Analysis}
    \label{sec:vae+pca}
    The observed degeneracy and entanglement among the VAE latent features limit their interpretability and practical usefulness, reflecting the difficulty of the VAE framework in establishing a direct correspondence between individual latent features and galaxy properties. Here, we  apply the PCA technique to transform the latent feature space into a set of orthogonal components (\S~\ref{sec:pca}) and further distill the dominant modes of variation that capture the most significant visual information. %ranked by their contribution to the overall variance. 
    This process recombines entangled features to align with the dominant directions of variance, thereby enhancing interpretability and enabling further dimensionality reduction by removing redundant components with negligible variance contributions.     
    
    \subsection{Optimal number of extracted features from machine vision}
    \label{sec:optimal_number_of_extracted_features}
        \begin{figure}
            \centering
            \includegraphics[width=\columnwidth]{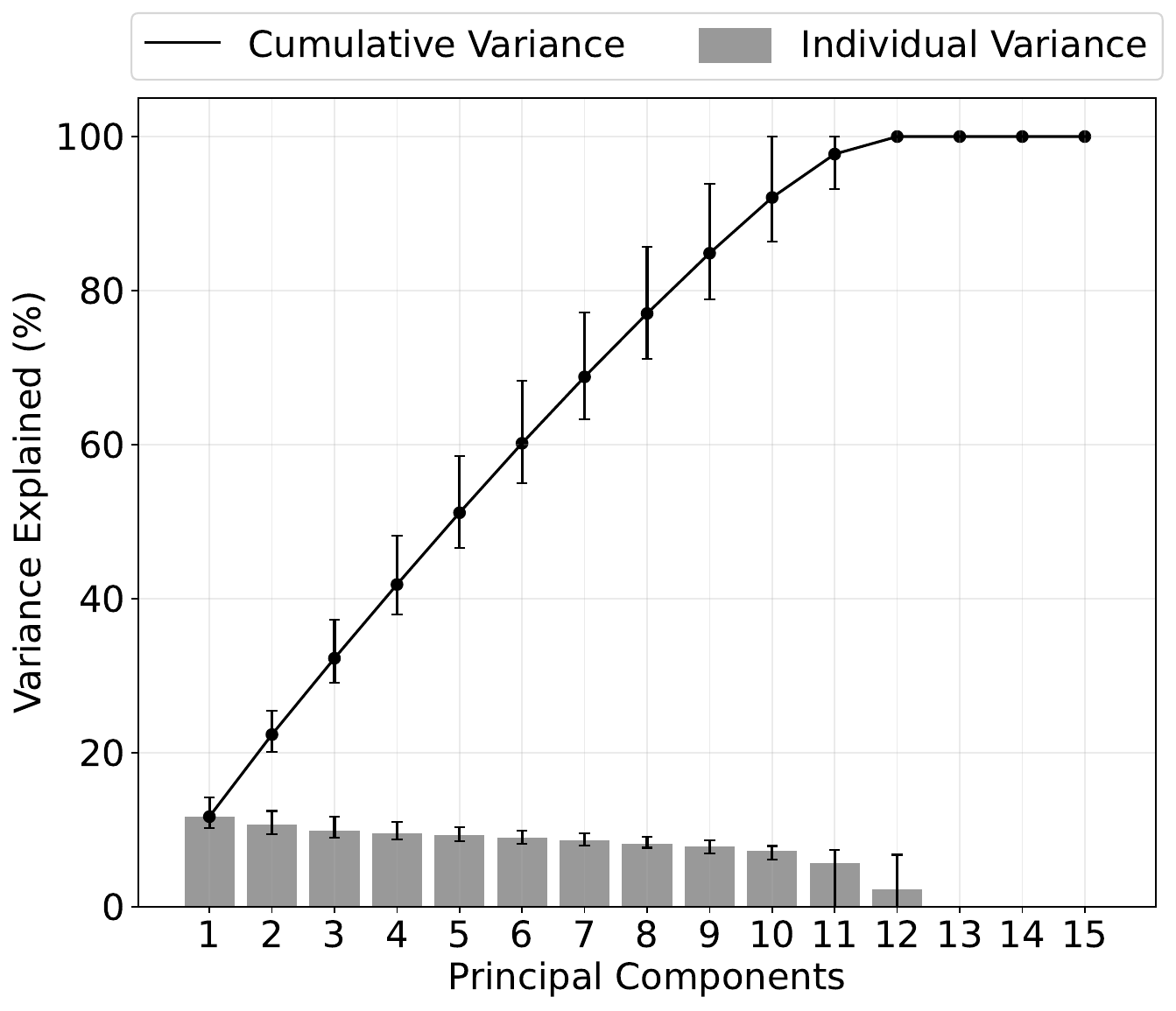}
            \caption{The PCA extracted features ranked by their explained variance. The index denotes the principal components ordered by the amount of variance each explains. Over 25 runs, the gray bars represent the mean variance explained by individual components, while the black solid line indicates the mean cumulative variance. Error bars show the minimum and maximum values across 25 runs.}
            \label{fig:pca_variance}
        \end{figure}

        One of our main goals is to investigate whether there exists an explicit number of meaningful features that can be extracted or learned from galaxy images. According to the bottom panel of Fig.~\ref{fig:loss_plots}, the number of PCs required to explain 99.9 per cent of the total variance reaches a plateau once at least 35 latent features are used in the VAE feature learning process. In Fig.~\ref{fig:pca_variance}, we further investigate the variance contributed by each PC and how the cumulative variance changes with the number of PCs. Based on 25 runs with 35 latent features, the results show that 10-12 PCs are sufficient to capture 99.9 per cent of the variance in the data. Each PC contributes approximately equally to the total variance ($\lessapprox10$ per cent), except in cases where 11 or 12 PCs are selected. In most runs, 11 PCs are sufficient to explain 99.9 per cent of the total variance, in approximately half of the cases. This provides a clear range for the number of features needed to capture the majority of the variance in the images. Thus, hereafter, we use the runs with 11 PCs as representative examples for discussion.
        
        However, it should be noted that this study is based on mock images from the EAGLE simulation; therefore, this number is expected to vary with image quality and the complexity of the objects presented in the images. In \S~\ref{sec:subset_training}, we examine how the complexity of the objects in images, as reflected in their morphologies, influences the extracted features and the determination of the optimal number of features to retain. In future work, we will investigate how this number changes when applied to real data from different surveys, in order to assess the extent to which attributes from galaxy images can be observed and interpreted through machine vision.
        
        \subsection{Comparison with structural measurements}
        \label{sec:structure_comparison}

        \begin{figure*}
            \centering
            \includegraphics[width=\textwidth]{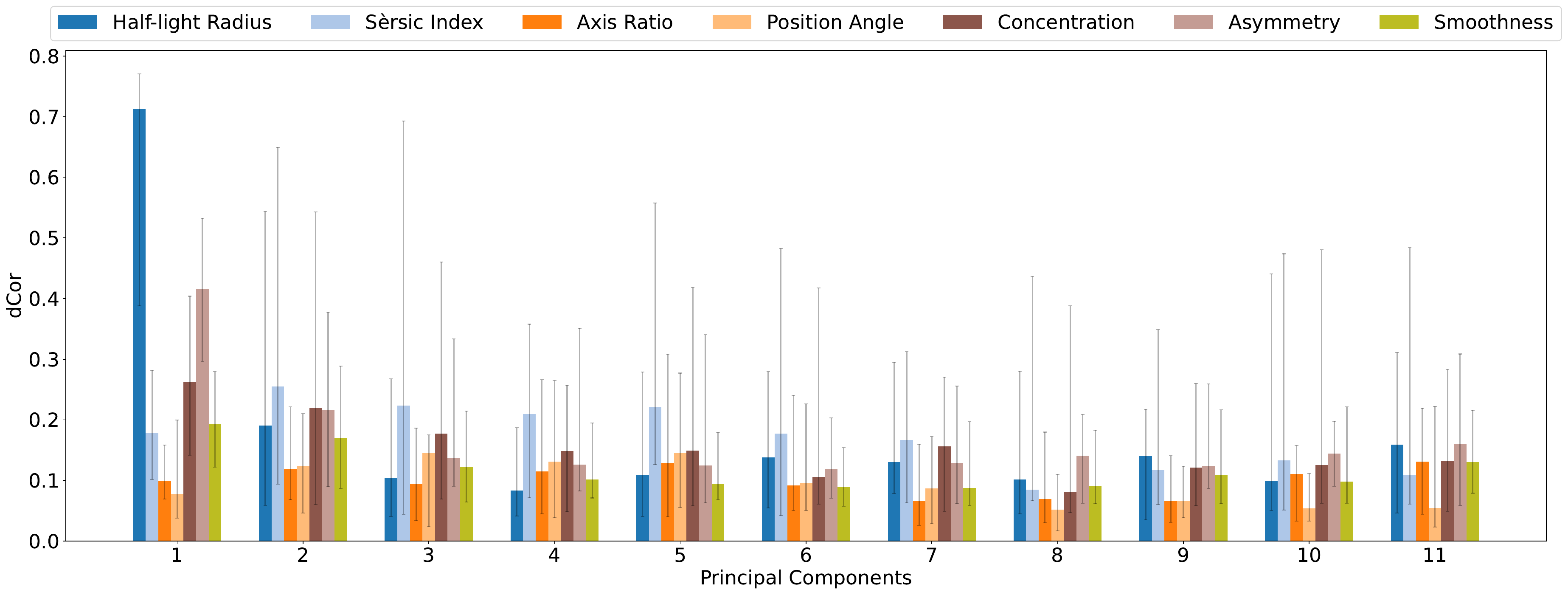}
            \caption{The distance correlations (dCor) between the principal components and structural measurements across runs acquired 11 extracted PCs. Different coloured bars represent different structural measurements. The height of the bars indicate the mean values across the runs, while the black vertical lines represent the uncertainty, quantified by the maximum and minimum values observed across the runs.}
            \label{fig:dcor_balanced}
        \end{figure*}

        \begin{figure*}
            \centering
            \includegraphics[width=\textwidth]{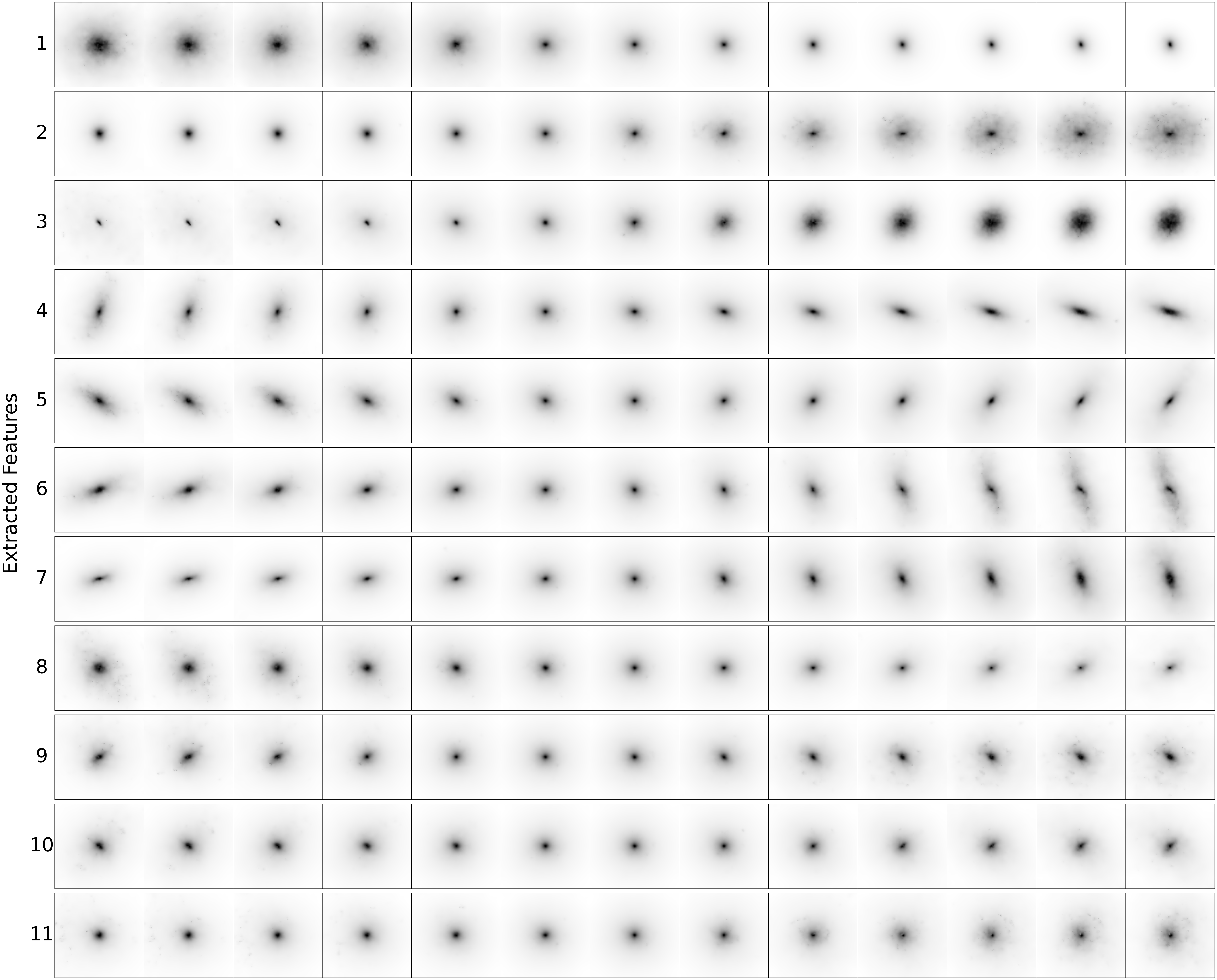}
            \caption{Example showing the effect of varying each PCA extracted feature from one of the runs acquiring 11 PCs. The index corresponds to each extracted feature, while all other features are fixed at their mean value except for the one specified.}
            \label{fig:transition_pca_features}
        \end{figure*}
        
        \begin{figure*}
            \centering
            \includegraphics[width=\textwidth]{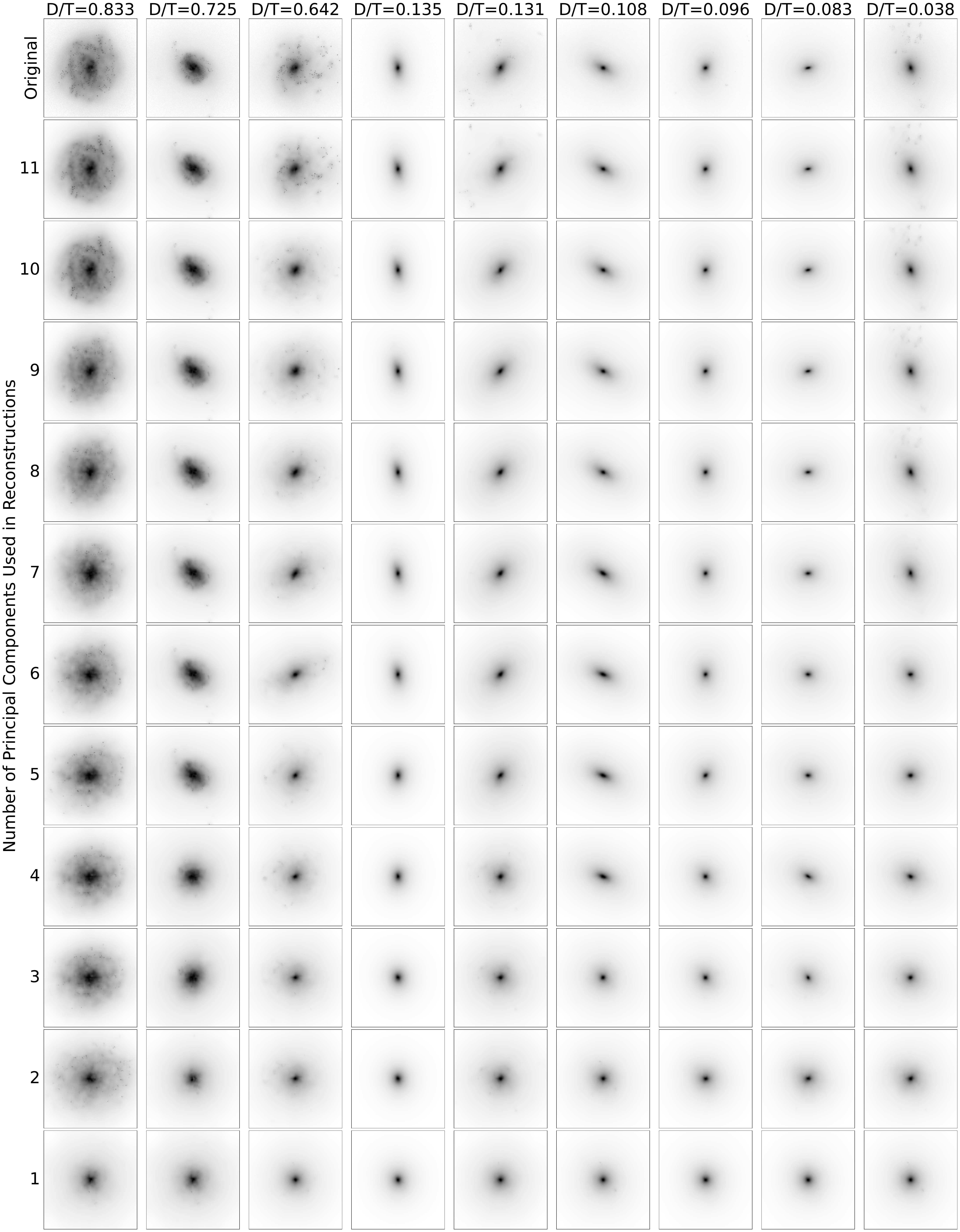}
            \caption{Example of reconstructions using PCA features. The top row shows the original images, with the $D/T$ ratio written as a label, while the rows below illustrate reconstructions using progressively fewer PCA features, with components contributing the lowest variance  gradually removed. The bottom row shows reconstructions using only the most dominant PC, which captures the largest variance in a single PC.}
            \label{fig:reconstruction_pca_cumulative}
        \end{figure*}
        
        Armed now with the linear combinations of the latent features resulting from the PCA, we again examine the correlation between the extracted features and the structural measurements (ie. the quantities taken from the references listed in \S~\ref{sec:vae_structures}). The entanglement between features is improved following the PCA process -- i.e. fewer components are correlated with the same structural measurements, and each component carries information relating to a smaller number of structural measurements. For example, compared to the cases of VAE latent features, where $14.5\pm1.0$ and $6.0\pm1.5$ features are correlated with the size of the galaxy and the S\'ersic index, respectively (\S~\ref{sec:vae_structures}), the PCA process extracts only approximately $2.0\pm1.0$ PCs that strongly correlate ($\mathrm{dCor}\gtrapprox0.3$) with either quantity. Moreover, Fig.~\ref{fig:dcor_balanced} shows that, across the runs retaining 11 PCs, the first PC, which contains the largest variance, dominates the correlation with the galaxy size (half-light radius). When examining the correlations on the basis of individual runs, the second PC occasionally shows a strong correlation, but generally is much weaker than the first PC. In the example shown in Fig.~\ref{fig:transition_pca_features}, the first and second PCs have $\mathrm{dCor}\sim0.62$ and $\sim0.47$ with the half-light radius of galaxies, respectively. This indicates that the size of galaxies generally contributes the most variance in image construction; in other words, the primary aspect the VAE framework prioritises during reconstruction is to accurately recover the size of galaxies.

        On the other hand, Fig.~\ref{fig:dcor_balanced} shows that the correlation with the S\'ersic index becomes moderately stronger between second and fifth PCs, with a peak at the third PC. This suggests that the description of internal disk/bulge structures emerges after the overall galaxy size is established during reconstruction. However, there is no strong indication of which specific PCs govern the generation of disk/bulge structures with stronger correlations. When examining the individual runs, we find that, in general, the formation of disk/bulge structures is governed by up to three extracted features per run, with $\mathrm{dCor \gtrapprox 0.3}$, though no specific components can be uniquely assigned to this process. Nevertheless, in cases where a dominant PC with a distance correlation of $\mathrm{dCor} > 0.6$ is present, it is typically the second or third component, which consequently drives the relatively higher mean dCor values with the S\'ersic index observed in Fig.~\ref{fig:dcor_balanced}. In Fig.~\ref{fig:transition_pca_features}, the third component in this example is strongly correlated with the S\'ersic index ($\mathrm{dCor\approx0.69}$), showing a clear transition from a concentrated bulge structure on the left to a diffuse disk structure on the right. This situation implies that, under certain conditions, the model can isolate the generation of disk/bulge structures, assigning them greater relative importance as they contribute significantly to the variance in the data (i.e. ranking second or third). However, our experiments show that, in general, disk and bulge structures are not captured by a single dominant component, but instead are encoded more diffusely across multiple correlated components under a VAE framework.
        
        For other structural measurements, such as the axis ratio and position angle, no single PC is clearly responsible for their variation, suggesting that these properties are distributed between multiple components. The 
        concentration shows slightly increased correlations with the top 3 PCs. The asymmetry exhibits a trend similar to galaxy size, with its dominant correlation occurring with the first PC. Smoothness also shows a decreasing trend similar to that of half-light radius and asymmetry, though with a smaller change between components. 
        
        Nevertheless, similar to the cases in \S~\ref{sec:vae_structures}, we note again that although the PCA process feature entanglement and improves interpretability by combining latent features and discarding redundant ones, the extracted features are still not directly comparable to structural measurements, as they still represent fractional contributions to the visual appearance of galaxies rather than exact structural properties. This exercise aims to provide insight into how machine vision interprets and reconstructs galaxy images, revealing how these representations relate to known structural properties and what kind of visual features are prioritised during reconstruction. In Fig.~\ref{fig:reconstruction_pca_cumulative}, we show reconstructed images using different numbers of extracted features. The first row shows the original images, followed by reconstructions with progressively fewer PCs. The bottom row illustrates reconstructions using only the first PC, which accounts for the largest fraction of variance in the data ($\gtrapprox10$ per cent) and, as discussed above, is primarily driven by galaxy size. In this example, the disk and bulge structures are generated when the third PC is included in the image reconstruction, while extended structures and asymmetries appear upon adding the fourth and fifth PCs, respectively. Finally, when 11 PCs are used, the reconstructions achieve an accuracy comparable to those obtained using all 35 latent features (see Fig.~\ref{fig:vae_example}).

    \section{Extracted features learned from different morphologies}
    \label{sec:subset_training}
    
        \begin{figure}
            \centering
            \includegraphics[width=0.6\columnwidth]{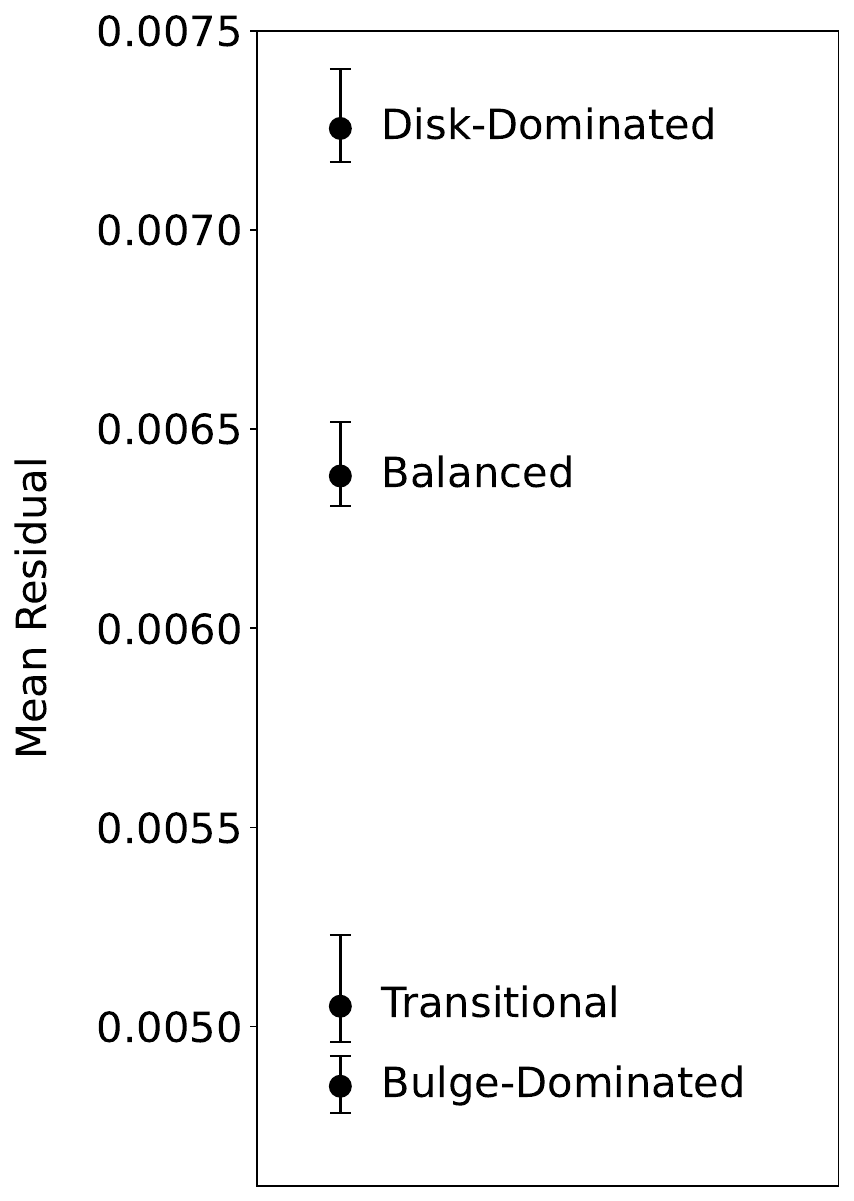}
            \caption{Reconstruction residuals between the input and reconstructed images for each subset and for the balanced mixed-morphology dataset. The points show the mean residual over 25 runs in which the initial weights of the neural networks are randomly varied in each run. The error bars connect the lowest and highest values of the residuals over these runs.}
            \label{fig:reconstruction_residuals}
        \end{figure}
        \begin{figure*}
            \centering
            \includegraphics[width=\textwidth]{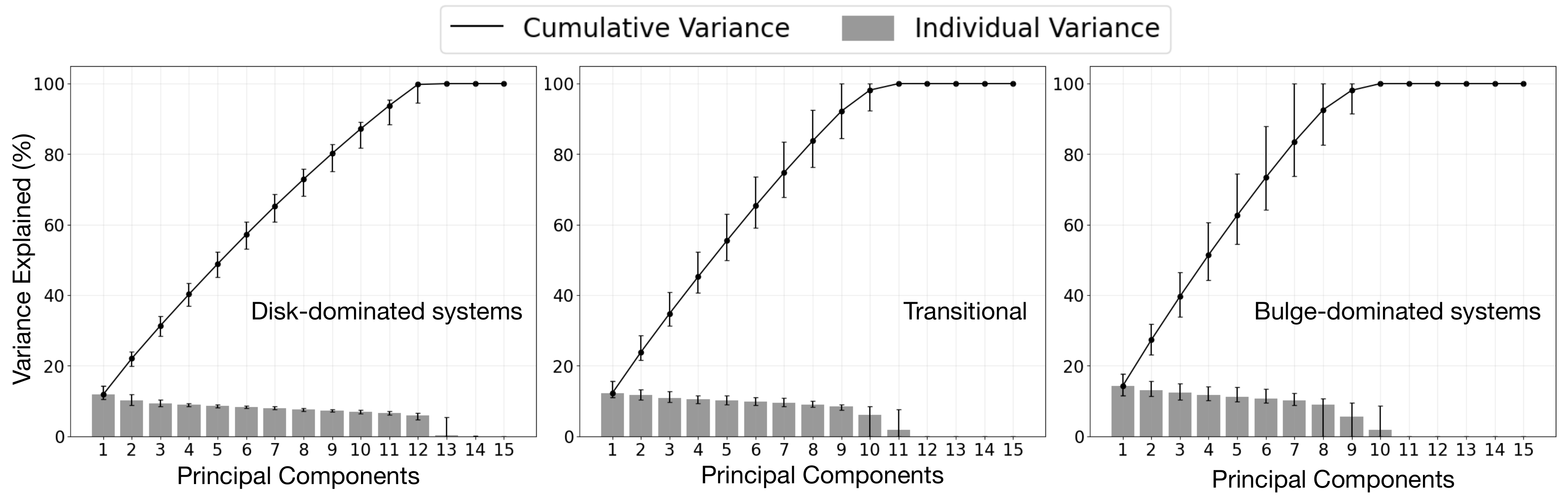}
            \caption{Same as Fig.~\ref{fig:pca_variance}, but for the VAE analysis and PCA applied to the subsamples defined by disc-to-total stell  three subset trainings. Each panel shows the results from 25 runs, presented from left to right for disk-dominated, transitional, and bulge-dominated systems, respectively, as labelled.}
            \label{fig:subset_variance}
        \end{figure*}

        \begin{figure*}
            \centering
            \includegraphics[width=\textwidth]{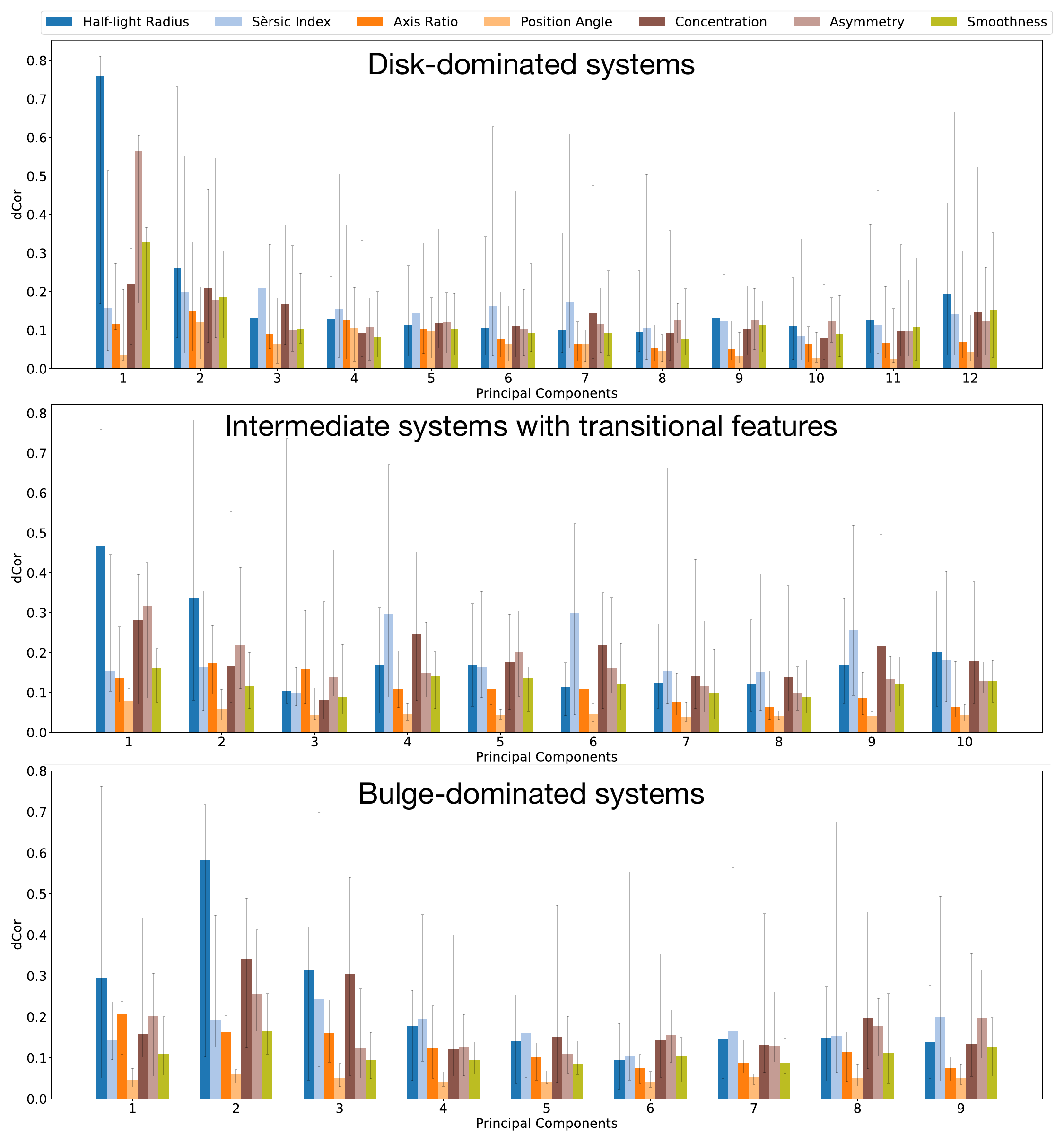}
            \caption{Same as Fig.~\ref{fig:dcor_balanced}, but for the three subset trainings. Each panel shows the results based on the runs obtained with the optimal number of PCs for the respective dataset.}
            \label{fig:dcor_subset}
        \end{figure*}

        The dataset used for the VAE analysis contains a mixture of morphologies. Consequently, the VAE balances the generation of diverse visual structures in an attempt to ensure consistent reconstruction performance across different morphological types. To investigate this further, we split the original data sample into three subsets with different morphologies, as determined by the disk-to-total stellar mass ratio ratio: disk-dominated systems ($D/T>0.2$), intermediate systems ($0.1\leq{D/T}\leq0.2$), and bulge-dominated systems ($D/T<0.1$). We repeat the VAE analysis for each subsample. We performed data augmentation to increase the total number of images in each subsample to be comparable in number to the original mixed-morphology dataset, thereby mitigating any potential biases introduced by differences in sample size. The subsamples were preprocessed in the same way as the main sample, following the description in \S~\ref{sec:pre-processing}. The purpose of this exercise is to examine whether the correlations between the extracted features and the structural measurements change when the model is trained exclusively on each morphological subset.

        Fig.~\ref{fig:reconstruction_residuals}  presents the comparison of the mean reconstruction residuals for each subset and the mixed-morphology dataset, computed as the pixel-wise difference between the input and reconstructed images. This shows that the disk-dominated galaxies, which contain more extended and asymmetric structures, generally exhibit higher reconstruction residuals compared to bulge-dominated systems and the mixed-morphology cases, which either intrinsically possess both characteristics of disk- and bulge-like systems (i.e. transitional) or from the mixed-morphology dataset. These results indicate that the reconstruction capability of the VAE framework is closely tied to the morphological complexity of galaxies. Moreover, in Fig.~\ref{fig:subset_variance}, we show the variance explained by each PC together with the cumulative variance across components. This demonstrates that the more complex the galaxy morphology, the larger the number of PCs required to contain 99.9 per cent of the total variance. In particular, disk-dominated galaxies, which as mentioned possess more extended, asymmetric, and in some cases exhibit spiral structures, require up to 12 PCs to capture over 99.9 per cent of the variance. Whereas, bulge-dominated systems and intermediate systems with transitional morphologies satisfy the same variance threshold with only 9 or 10 PCs, respectively. When compared with the result from the original mixed $D/T$ dataset, which draws 11 PCs, the optimal number of PCs required follows the same order as in Fig.~\ref{fig:reconstruction_residuals}, indicating that the optimal number of components scales with the morphological complexity of the galaxies in the images.

        In Fig.~\ref{fig:dcor_subset}, we further examine the distance correlation (dCor) between the extracted features and structural measurements for the three morphological subsets using the runs with the optimal number of PCs for each respective subset. Overall, the correlations with galaxy size are consistent with the results presented in \S~\ref{sec:structure_comparison}: galaxy size shows strong correlations with the first and/or second PCs, indicating that size is the primary feature the VAE prioritises during reconstruction. A subtle difference is seen between different morphologies: disk-dominated galaxies show strong correlation predominantly with the first PC, whereas bulge-dominated systems show a peak at the second PC. Galaxies with transitional morphologies then distribute their correlations between the first and second PCs with an emphasis on the first component. On the other hand, consistent with the discussion in \S~\ref{sec:structure_comparison}, the correlations with S\'ersic index are distributed diffusely across components, and it remains unclear whether a single dominant PC governs the reconstruction of disk and bulge structures. Specifically, the correlation trend is weak for disk-dominated galaxies, slightly stronger for bulge-dominated galaxies with a peak at third component, and more pronounced for systems with transitional morphologies, where three relatively dominant features appear at fourth, sixth, and ninth PC. This suggests  that the VAE framework requires further effort to accurately differentiate and reconstruct disk and bulge structures in galaxies with transitional morphologies, where both features coexist, compared to disk- or bulge-dominated cases.

        Axis ratio and position angle follow the same trend described in \S~\ref{sec:structure_comparison}; the features contributing to these properties are distributed across all components. However, we notice a minor emphasis on position angle for disk-dominated galaxies, and a slight increase in the dominance of axis ratio correlations among the top three PCs for bulge-dominated systems. Concentration presents more prominent correlations in bulge-dominated systems compared to the other two types, whereas smoothness and particularly asymmetry show stronger correlations with features from disk-dominated galaxies.
    
        This experiment demonstrates that the complexity of objects in the images not only influences the optimal number of extracted features, but also affects the relative emphasis of different components of galaxy appearance during reconstruction. A summary of the individual emphasis on different structures across morphologies is provided in the Conclusions (\S~\ref{sec:conclusion}).

    \section{Machine vision and physical properties of galaxies}
    \label{sec:physical_comparison}
        \begin{figure*}
            \centering
            \includegraphics[width=\textwidth]{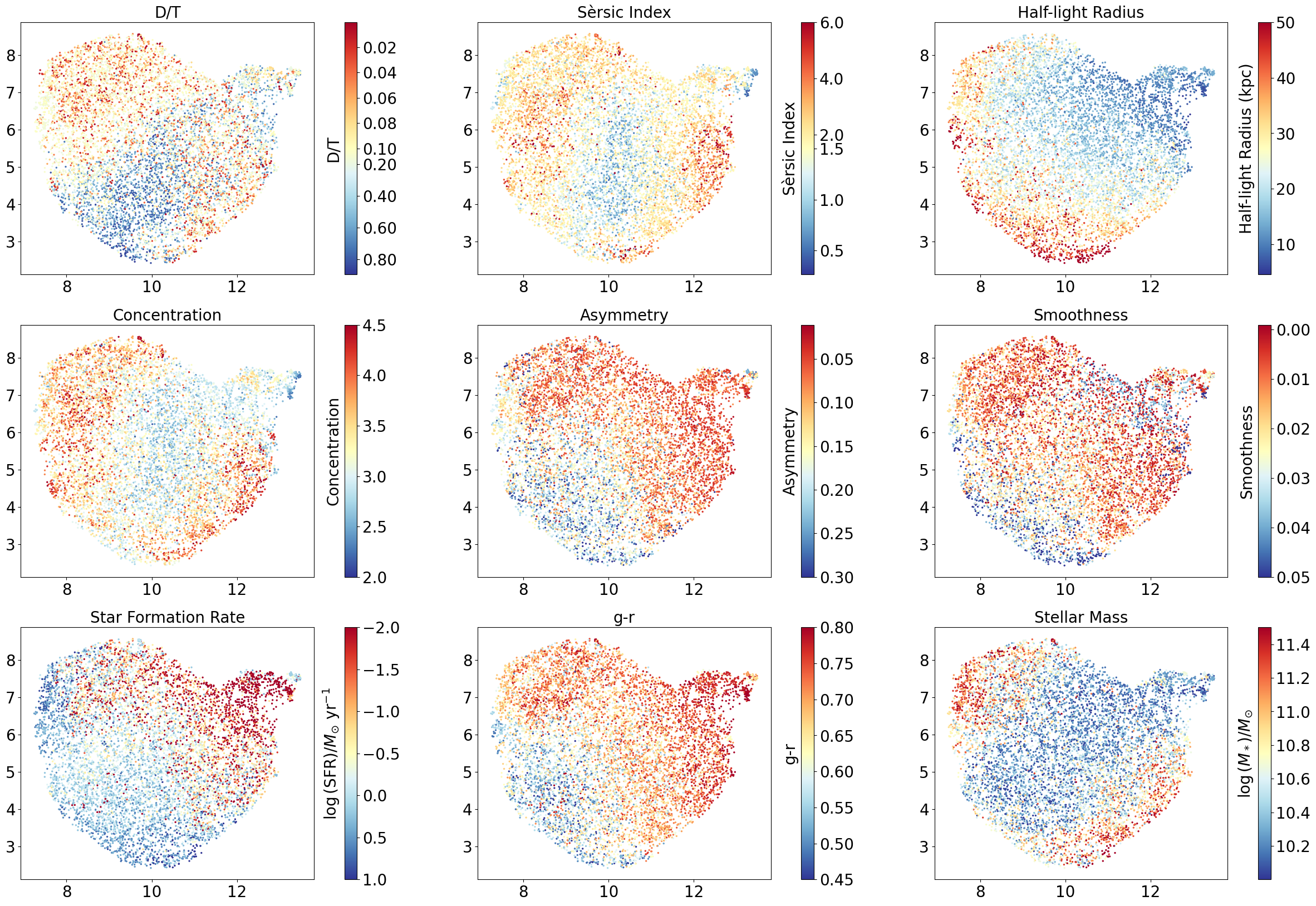}
            \caption{UMAP visualisations of the extracted PCA features. The colour palette in each panel represents structural measurements (\textit{top:} D/T, S\'ersic index, half-light radius; \textit{middle:} concentration, asymmetry, smoothness) and physical properties of galaxies (\textit{bottom:} star formation rate, colour $g-r$, stellar mass). The D/T and S\'ersic index color bars are adjusted to correspond to the typical ranges of the defined morphological types here. Note that some of the colour bars are inverted with the higher value of the quantity coloured blue rather than red (star formation rate, $D/T$, asymmetry, smoothness).}
            \label{fig:umap_property}
        \end{figure*}
    
        As discussed above, each extracted feature contributes fractionally to the overall image and the visual appearance of galaxies. Therefore, it is necessary to employ summary statistics that integrate all extracted features within the high-dimensional feature space, enabling each galaxy to be represented holistically, providing a machine-vision perspective, and facilitating the examination of its connections with galaxy properties. 
        
        For this purpose, we apply Uniform Manifold Approximation and Projection \citep[UMAP;][]{McInnes2018,McInnes2018_software}, a non-linear dimensionality reduction technique that projects galaxies from the high-dimensional feature space, defined by the extracted PCA features, into a low-dimensional representation. UMAP constructs a graph by computing distances between all objects in the high-dimensional feature space and establishes clusters based on nearest neighbours. The construction of the graph is an iterative procedure. The initial guess at the graph is  made by performing a `spectral embedding', which results in a distribution of points in the space that is similar to a self-avoiding pattern. Adjustments are made to this starting point by minimising a cross-entropy loss function\footnote{A related technique to UMAP is the t-SNE method of \cite{VanDerMaaten2008}. The t-SNE starts from a random initialisation and  so can take longer to converge than UMAP. Also, the optimisation funcion used is different in t-SNE than in UMAP.}. UMAP iteratively adjusts the positions of objects in the low-dimensional space to preserve neighbourhood relationships from the high-dimensional space. The UMAP approach preserves both local and global relationships, providing an informative low-dimensional representation of the high-dimensional feature space.

        %UMAP enables us to represent each galaxy holistically in low-dimensional space and directly compare these representations with the structural measurements and physical properties of galaxies. 
        Fig.~\ref{fig:umap_property} shows the UMAP visualisations of our balanced mixed-morphology dataset (\S~\ref{sec:data_augmentation}). Firstly, galaxies with different characteristics are located at distinct positions in these maps, indicating that machine vision perceives them as different. In a clustering context, these galaxies would naturally be assigned to separated groups.
        
        The top-left panel ($D/T$) of Fig.~\ref{fig:umap_property} presents the distribution of disk fractions (which is used as an indicator of galaxy morphology here) in the UMAP space. This shows that, broadly speaking, machine vision recognises three distinct groups from top-left to bottom-right in UMAP space, corresponding to different combinations of disk and bulge components. Bulge-dominated/intermediate galaxies (red/yellow) occupy two separate regions in the map, suggesting that while they share similar bulge fractions, machine vision distinguishes them as visually different populations. The S\'ersic index shows a similar trend, though with different distributions. This discrepancy happens because the $D/T$ values we use here are directly derived from kinematic data in simulations, while the S\'ersic index is measured from mock SDSS-quality images. Additionally, the criteria used to define morphology classes in both cases are not strictly consistent with one another. In the following discussion, we use the $D/T$ distributions for indicating the genuine disk and bulge fractions of galaxies. 

        Overall, we observe two distinct trends in these maps: one follows the $D/T$ distributions, such as S\'ersic index, concentration, and stellar mass, broadly separating the map into three regions; while the other splits the map into two areas, located at the bottom-left and top-right, like half-light radius, asymmetry, smoothness, star formation rate (SFR), and galaxy colour $g-r$. Based on these two trends, we can roughly categorise the samples into 6 approximate classes, each characterised by the following properties:
        \begin{enumerate}
            \item bulge-like systems (low $D/T$, high S\'ersic index), smaller size, high concentration, low asymmetry, low smoothness, likely quenched, redder in colour, and lower stellar mass (e.g., samples located around [9, 7.5]); 
            \item bulge-like systems, (lower $D/T$, high S\'ersic index), larger size, high concentration, high asymmetry, high smoothness, likely star-forming, and higher stellar mass (e.g., samples located around [8, 7]); 
            \item disk-like systems (high $D/T$, low S\'ersic index), smaller size, low concentration, low to intermediate asymmetry, low smoothness, intermediate star forming activities with intermediate $g-r$ colour, low stellar mass (e.g., samples located around [10, 6]); 
            \item disk-like systems (high $D/T$, low S\'ersic index), larger size, intermediate concentration, high asymmetry, high smoothness, likely star-forming, lower stellar mass (e.g., samples located around [10, 4]); 
            \item intermediate systems (mixed $D/T$, intermediate S\'ersic index), small size, low concentration, low asymmetry, mixed smoothness, quenched, low stellar mass (e.g., samples located around [12, 6.5]); 
            \item intermediate/bulge-like systems (mixed $D/T$, intermediate to high S\'ersic index) intermediate size, high concentration, low asymmetry, low smoothness, weak star formation activities with redder colour, larger stellar mass (e.g., samples located around [12, 4]).
        \end{enumerate}

        \noindent Note that this categorisation is approximate and is used here solely to facilitate discussion. A proper clustering analysis could be performed to more rigorously investigate the machine-defined galaxy classes as a machine-learning version of the Hubble sequence. However, we consider this analysis to be beyond the scope of the current paper.

        This approximate territorial division of the UMAP space serves as a broad-bush guideline for distinguishing galaxy populations, that (i) resemble typical bulge-dominated systems; (ii) are star-forming, clumpy, and asymmetric galaxies that somehow contain a bulge component; (iii) are likely to be so-called `green-valley' galaxies that are low-mass, small, disky and diffuse systems without significant clumpiness or asymmetry, exhibiting intermediate levels of star formation; (iv) resemble typical star-forming disk galaxies; (v) are diffuse, symmetric, quenched, low-mass galaxies with mixed disk and bulge structures; and (vi) appear likely to  represent systems transitioning toward typical quenched elliptical galaxies. This demonstrates that machine vision, which captures visual features from images, delivers more detailed information than conventional structural measurements or kinematic indicators like $D/T$. Except for stellar mass, which broadly follows the $D/T$ trend, disk and bulge structures are not unique indicators of underlying physical properties. Although combining multiple structural measurements -- for instance, S\'ersic index and asymmetry -- can improve the indication of underlying physical properties, such approaches remain limited in their ability to separate galaxy populations. As shown here, and in agreement with earlier results, different structural measurements do not correspond perfectly to the underlying physical properties. We demonstrated that machine vision, however, is capable of distinguishing subtle variations in visual morphology from images. Also, consistent with the findings of C21, machine vision can identify systems that share similar structural characteristics yet differ in their physical properties, thereby representing distinct populations of galaxies.

        %In order to consider all components, one can apply clustering algorithm to conduct unsupervised categorsiation.... 
    
        %The physical properties of galaxies provide direct insight into the processes that have formed and transformed the galaxies, and are therefore closely linked with their structure and visual morphologies. Existing structural measurements and visual morphologies have demonstrated correlations with physical properties. For instance, spiral galaxies typically exhibit active star formation activities and appear bluer in optical colours, whereas elliptical galaxies tend to be more passive, hosting older stellar populations, and displaying redder colours. Since the vast morphological classification available from Galaxy Zoo projects \citep{Lintott2008,Lintott2011}, we have begun to identify exceptions to this relationship, such as red spirals with little star formation or blue ellipticals that show signs of recent starburst activity. These exceptions indicate a potential deficiency in existing structural and morphological schemes. As mentioned in Section~\ref{sec:intro}, C21's unsupervised machine classification scheme found correlations with galaxy colours using monochromatic images, suggesting a potentially subtle yet direct link between visual morphologies captured in galaxy images and their underlying physical properties. In this work, we examine if any of our extracted features also capture similar trend.

    \section{Discussion and Conclusions}
    \label{sec:conclusion}

    We have used mock galaxy images generated from the EAGLE simulation by \cite{Trayford2017} to investigate the features learned by machine vision through a widely used self-supervised technique,  the variational autoencoder (VAE) framework. We compress galaxy images of size $256 \times 256$ pixels into a latent space of 35 features, a number determined by balancing both the reconstruction residuals and the KL divergence terms. Then, we apply principal component analysis (PCA) to recombine the latent features and rank them according to their contributions to the total variance.
    
    The learned latent features focus on different regions of the images depending on their morphologies -- from the outskirts in disk-dominated galaxies to the inner regions in bulge-dominated systems. However, while correlations can be observed between the latent features and structural measurements, it is important to recognise that these representations are not directly equivalent to structural measurements that require a predefined mathematical form. Instead, the latent features, along with the sequentially extracted principal components (PCs), serve as data-driven descriptors of the visual appearance of galaxies, with each feature encoding fractional contributions to the overall morphology. However, by incorporating a PCA, we reduce the entanglement between learned features and structural measurements. For example, the number of features correlated with half-light radius and S\'ersic index decreases from $14.5\pm1.0$ and $6.0\pm1.5$ VAE latent features, respectively, to only $2.0\pm1.0$ PCs after the PCA process. 
    
    We also find that the VAE framework prioritises the reconstruction of galaxy size, as evidenced by the strong correlation of this quantity with the highest-ranked PCs (see Fig.~\ref{fig:dcor_balanced} and Fig.~\ref{fig:dcor_subset}). While one may consider galaxy size redundant as it is not a primary factor in visual classifications of galaxy morphology, Fig.~\ref{fig:umap_property} shows that galaxies with similar D/T values can have different size, as well as different physical properties such as SFR, colour and stellar mass. This provides a new insight into this line of study. Once the size is established, the framework subsequently incorporates internal structures such as disk and bulge components. By examining the correlation between extracted PCs and the S\'ersic index, we find that, in general, these structures are not governed by a single component but are instead distributed across multiple PCs. However, for intermediate systems with transitional morphologies, where both structures coexist, the generation of disk and bulge structures becomes more dedicated compared to purely disk- or bulge-dominated cases. We also find that asymmetric structures are more emphasized in disk-bulge galaxies, showing strong correlations with the first PC, whereas concentration is relatively more prominent in bulge-dominated systems. Additionally, a minor emphasis on position angle is observed in disk-dominated galaxies, whereas the axis ratio shows slight prominence in bulge-dominated systems. 
    
    We summarise the representation of the extracted PCs in the VAE reconstruction for galaxies of different morphologies as follows: 
    \begin{itemize}
        \item \textbf{Disk-dominated galaxies} primarily establish the overall galaxy size and asymmetric structures, with an additional but minor emphasis on smoothness. The reconstruction of the position angle shows a slight enhancement at the second PC, compared to the correlations of other PCs. While no single PC strongly dominates the S\'ersic index, a noticeable peak occurs at the third PC.
        \item \textbf{Bulge-dominated galaxies} also primarily capture the overall size of the galaxy. However, unlike disk-dominated galaxies, this is distributed across the top three PCs, with a peak at the second PC. The S\'ersic index follows a similar trend to that of disk-dominated cases, with a peak at the third component. Then, compared to disk-dominated cases, these galaxies show weaker correlations with asymmetric structures with a lower peak at the second PC. In contrast, a greater emphasis is shown on concentration compared to disk-dominated cases, with higher correlations observed at the second and third PCs. Additionally, no noticeable correlation with position angle is found, while a slight emphasis on axis ratio is shown at the first PC with decreasing correlations across subsequent components.
        \item \textbf{Intermediate galaxies with transitional morphologies} exhibit a mixture of correlations between the other two types. The overall galaxy size remains the primary characteristic in reconstruction, accompanied by concentration and asymmetry. Noteably, the correlation with the S\'ersic index becomes more prominent in this subset, displaying three clear peaks, which indicates greater focus on accurately generating disk and bulge structures in this subset. 
    \end{itemize}

    Furthermore, from Fig.~\ref{fig:loss_plots} and Fig.~\ref{fig:pca_variance}, we conclude that, through machine vision using the VAE framework, there exists an optimal range of approximately 10-12 meaningful features that can be extracted from galaxy images and capture over 99.9 per cent of the variance in our dataset. Nevertheless, we also find that it is essential to have at least 35 latent features in the VAE process to ensure sufficient information has been learned for the PCA step, enabling the extraction of this optimal number of components. This number can vary depending on image quality and the intrinsic complexity of the galaxies, but our analysis demonstrates that an optimal range is achievable and can be systematically investigated.

    Finally, we project each sample from the high-dimensional feature space into a two-dimensional representation using the UMAP technique. Galaxies at different positions on this map possess different visual features identified by machine vision. Certain structural properties are known to be associated with specific populations; however, exceptions do exist. By comparing their positions on the map with structural measurements and physical properties, we can quickly and easily identify these outlier populations whose structures deviate from expectations. This approach demonstrates that these galaxies possess distinct visual features which may not be apparent through structural measurements or human classification, yet can be directly and effectively distinguished from common cases by machine vision.
    
    We investigated the use of the VAE framework combined with PCA for extracting visual features from images. We examined and discussed the optimal number of features that can be acquired from galaxy images, their representation in images, and their correlations with known structural measurements. This provides a foundation for efficient data compression and feature extraction. By linking these features to physical properties, the approach supports the physical interpretation of galaxy images which may be used to examine observational images in the future. This demonstrates the effectiveness of machine vision in categorising both common galaxy types and populations with atypical structures, offering a direct way to distinguish galaxy populations. Overall, the results show that machine-learned features provide a data-driven representation of galaxy morphology, complementing traditional structural measurements and revealing new visual and physical characteristics. We hope that the results presented here will guide improvements in machine learning methods for feature extraction and image reconstruction, while also supporting tasks such as classification, clustering, and anomaly detection.

\section*{Acknowledgements}
We thank Lucas Bignone for providing a catalogue of structural measurements and Ryan Cooke for helpful discussions. TYC acknowledges support from the Science Technology Facilities Council (STFC) through ST/T000244/1. CMB acknowledges support from STFC via ST/X001075/1.  
%The GPUs resources are NVIDIA V100 (gn001) and A100 (mad06), mounted on the COSmology MAchine (COSMA).
This work used the DiRAC@Durham facility managed by the Institute for Computational Cosmology on behalf of the STFC DiRAC HPC Facility (www.dirac.ac.uk). The equipment was funded by BEIS capital funding via STFC capital grants ST/K00042X/1, ST/P002293/1, ST/R002371/1 and ST/S002502/1, Durham University and STFC operations grant ST/R000832/1. DiRAC is part of the National e-Infrastructure.

%%%%%%%%%%%%%%%%%%%%%%%%%%%%%%%%%%%%%%%%%%%%%%%%%%
\section*{Data Availability}
The simulation data are publicly available through the EAGLE database: \url{https://eagle.strw.leidenuniv.nl/wordpress/index.php/eagle-simulations-public-database/}. The catalogues of structural measurements are mostly publicly available and can be retrieved from the corresponding papers, except for the catalogue from \citet{Bignone2020}, which we obtained on request.

%%%%%%%%%%%%%%%%%%%% REFERENCES %%%%%%%%%%%%%%%%%%

% The best way to enter references is to use BibTeX:

\bibliographystyle{mnras}
\bibliography{manuscript} % if your bibtex file is called example.bib

% Alternatively you could enter them by hand, like this:
% This method is tedious and prone to error if you have lots of references
%\begin{thebibliography}{99}
%\bibitem[\protect\citeauthoryear{Author}{2012}]{Author2012}
%Author A.~N., 2013, Journal of Improbable Astronomy, 1, 1
%\bibitem[\protect\citeauthoryear{Others}{2013}]{Others2013}
%Others S., 2012, Journal of Interesting Stuff, 17, 198
%\end{thebibliography}

%%%%%%%%%%%%%%%%%%%%%%%%%%%%%%%%%%%%%%%%%%%%%%%%%%

%%%%%%%%%%%%%%%%% APPENDICES %%%%%%%%%%%%%%%%%%%%%

%\appendix

%%%%%%%%%%%%%%%%%%%%%%%%%%%%%%%%%%%%%%%%%%%%%%%%%%

% Don't change these lines
\bsp	% typesetting comment
\label{lastpage}
\end{document}